\documentclass[journal]{IEEEtran}
\usepackage{times,amsmath,epsfig,latexsym,amssymb,psfrag}
\usepackage{cite}
\usepackage{amsthm}
\usepackage{algorithm}
\usepackage{algorithmic}
 \usepackage{multirow}
\usepackage{graphics, color, psfrag,amsfonts}
\usepackage{mathtools}
\usepackage{dsfont}
\usepackage{amssymb}
\usepackage{pifont}
\usepackage{tikz}
\usetikzlibrary{shapes,arrows}
\usetikzlibrary{decorations.markings}
\usetikzlibrary{shapes,arrows}
\usetikzlibrary{positioning}
\usetikzlibrary{decorations.pathreplacing}
\newcommand{\cmark}{\ding{51}}%
\newcommand{\xmark}{\ding{55}}%
\theoremstyle{definition}

 \newtheorem{remark}{Remark}
 \newtheorem{exmp}{Example}

\begin{document}
\ifCLASSOPTIONonecolumn
\title{\fontsize{23}{22}\selectfont  Random Linear Network Coding for Wireless Layered Video Broadcast: General Design Methods for Adaptive Feedback-free Transmission}
\else
\title{Random Linear Network Coding for Wireless Layered Video Broadcast: General Design Methods for Adaptive Feedback-free Transmission}
\fi
 \author{
   \IEEEauthorblockN{}
}
 \author{
  \IEEEauthorblockN{Mohammad Esmaeilzadeh,~\IEEEmembership{Student Member,~IEEE,} Parastoo Sadeghi,~\IEEEmembership{Senior Member,~IEEE,} and Neda Aboutorab,~\IEEEmembership{Member,~IEEE}
\thanks{The authors are with the Research School of Information Sciences and Engineering, The Australian National University, Canberra, 0200, ACT, Australia
(e-mail: \{mohammad.esmaeilzadeh, neda.aboutorab, parastoo.sadeghi\}@anu.edu.au).}
}
\thanks{This work was supported under the Australian Research Council Discovery Projects and Linkage Projects funding schemes (project nos. DP120100160 and LP100100588).}
}
\maketitle


\begin{abstract}

This paper studies the problem of broadcasting layered video streams over heterogeneous single-hop wireless networks using feedback-free random linear network coding (RLNC). We combine RLNC with unequal error protection (UEP) and our main purpose is twofold. First, to systematically investigate the benefits of UEP+RLNC layered approach in servicing users with different reception capabilities. Second, to study the effect of not using feedback, by comparing feedback-free schemes with idealistic full-feedback schemes. To these ends, we study `expected percentage of decoded frames' as a key content-independent performance metric and propose a general framework for calculation of this metric, which can highlight the effect of key system, video and channel parameters. We study the effect of number of layers and propose a scheme that selects the optimum number of layers adaptively to achieve the highest performance. Assessing the proposed schemes with real H.264 test streams, the trade-offs among the users' performances are discussed and the gain of adaptive selection of number of layers to improve the trade-offs is shown. Furthermore, it is observed that the performance gap between the proposed feedback-free scheme and the idealistic scheme is very small and the adaptive selection of number of video layers further closes the gap. 

\end{abstract}

\begin{IEEEkeywords}
Random Linear Network coding, Layered Video Streaming, Wireless Broadcast
\end{IEEEkeywords}

\section{Introduction}

\ifCLASSOPTIONtwocolumn
\IEEEPARstart{S}{ince}
\else
Since
\fi
the introduction of network coding (NC) in~\cite{Ahlswede:2000:IEEE-IT:NIF}, this technique has gained much attention and popularity in different areas of wired and wireless communications. Thanks to its capability in improving bandwidth utilization and reducing transmission delay and energy, NC has fitted well into different applications, from file and media transfer to sensor networks and also distributed storage systems~\cite{Chou:2007:IEEE-SPM-NCI}. However, with the constant demand for better quality of service in such applications and the consequent technology growth, new challenges in NC research are still emerging. 
One area that is particularly challenging is NC for video streaming~\cite{Magli:2013:IEEE-TMM:NCM}. 

In video streaming, timely delivery of reliable and high quality content is desirable, but this is often hindered by delay, packet loss and bandwidth limitations. These challenges are even more restrictive when video is transmitted over wireless networks. To deal with these challenges, a number of useful features has been added to video streaming standards. For instance, the scalable video coding (SVC) of H.264~\cite{Schwarz:2007:IEEE:SCV} provides layered video streams with various levels of quality, which can be useful when heterogeneity in users' reception capabilities or displays exists. While the added features can mitigate the video streaming challenges to some extent, combining the layered approach with forward error correction (FEC) techniques has shown to be even more beneficial~\cite{Huo:2015:IEEE:AL-FEC}, as it provides unequal error protection (UEP) for different importance layers and the quality of video streaming can be further improved. Some examples of such approach are studied for Reed-Solomon and Fountain codes in~\cite{Ha:2008:LUE,Vukobratovic:2009:IEEE-TMM:SVM}, respectively.

In this paper, we focus on random linear NC (RLNC) as a rateless FEC technique. The selection of RLNC is justified by its superior capability of simple extension to general networks (allowing re-encoding at intermediate nodes) that end-to-end FEC techniques (e.g., LT~\cite{Luby:2002:LTC} and Raptor~\cite{Shokrollahi:2006:IEEE-IT:Raptor} codes) do not share~\cite{Lucani:2012:IEEE-IT:NCD}. Furthermore, RLNC can provide better trade-offs among bandwidth efficiency, complexity and delay, compared to other FEC techniques as noted in~\cite{Magli:2013:IEEE-TMM:NCM}.

As one of the early works in the context of NC for video streaming, we can refer to~\cite{Hulya:2009:IEEE-JSAC:VAO}, where video-aware opportunistic NC over wireless networks is proposed. In this study, the importance of video packets is first determined based on the deadline and contribution to the video quality. Then, considering the decodability of packets by several users, efficient network codes to maximize the overall video quality are selected. Assessing the proposed scheme for different network topologies, significant gain of the video-aware opportunistic NC over scheduling algorithms without NC is shown. Another research that similarly considers quality and deadline of video packets is conducted in~\cite{Nguyen:2011:IEEE-TVT:JNC}. In this study, the authors consider layered video streams and propose to use the finite horizon Markov decision process (MDP) to select efficient network codes, not only by considering the next transmission, but also by taking into account all the transmissions before packets' deadline. Their scheme shows to outperform non-NC schemes in multiuser single-hop wireless networks for both broadcast and multiple unicasts scenarios.

In~\cite{Hulya:2009:IEEE-JSAC:VAO} and~\cite{Nguyen:2011:IEEE-TVT:JNC}, which are discussed above, XOR-based NC is used. While this type of NC has many favorable characteristics, the dependency of code selection on packet delivery acknowledgments (feedback) makes it unsuitable for some systems/networks such as large-scale broadcast networks or large-latency networks. Hence, RLNC that has lower dependency on feedback is studied for video streaming~\cite{Nguyen:2010:VSN,Thomos:2011:IEEE-TMM:PDV, Vukobratovic:2012:IEEE-TCOM:UEP, Nazir:2012:ALS:h264,Tassi:2015:IEEE-JSAC}. In these studies, the authors utilize layered video and propose to combine UEP with feedback-free RLNC (referred to as `UEP+RLNC' in this paper) to achieve an improved performance over non-NC schemes. 
In particular,~\cite{Nguyen:2010:VSN} proposes an RLNC framework for video streaming in content delivery networks (CDNs) and peer-to-peer (P2P) networks, where multiple servers/peers are employed to stream a video to a single user. The authors in~\cite{Thomos:2011:IEEE-TMM:PDV} consider distributed video delivery over lossy overlay networks,~\cite{Vukobratovic:2012:IEEE-TCOM:UEP, Nazir:2012:ALS:h264} study video streaming using RLNC for single receiver settings and in a recent study~\cite{Tassi:2015:IEEE-JSAC}, joint optimization of RLNC and resource-allocation methods for video streaming over generic cellular systems is investigated.

In~\cite{Nguyen:2010:VSN, Vukobratovic:2012:IEEE-TCOM:UEP}, the idea of coding across layers (i.e., inter-layer coding), as one step forward compared to coding only within layers (i.e., intra-layer coding) is proposed. In this method, which is referred to as hierarchical NC in~\cite{Nguyen:2010:VSN}, multiple expanding windows (EW)~\cite{Vukobratovic:2012:IEEE-TCOM:UEP} are proposed to generate RLNC packets. Then, using a probabilistic approach for selecting coding windows, the decoding probabilities of different layers of video are obtained and the gain of inter-layer UEP+RLNC over intra-layer UEP+RLNC is shown. Later, a similar idea is applied for real video streams encoded by H.264/AVC (single layer version of H.264/SVC) in~\cite{Nazir:2012:ALS:h264}. The authors manually generate importance layers based on the contribution of packets to the overall peak signal-to-noise ratio (PSNR) and also their deadline, and investigate how selecting coded packets from different layers affects the PSNR performance. The authors in~\cite{Tassi:2015:IEEE-JSAC} consider both intra- and inter-layer UEP+RLNC for Point-to-Multipoint services over multiple orthogonal broadcast erasure subchannels. They formulate packet error probability expression and incorporate it into their resource allocation frameworks to investigate the advantage of layered NC over multirate transmission. They adapt their framework to 3GPP Long Term Evolution-Advanced (LTE-A) standard and demonstrate the improvement in the quality of received H.264/SVC video.
 
While the aforementioned studies shed some light on the applicability of EW UEP+RLNC for streaming of layered video, a systematic study of this approach for multi-user broadcast of live H.264/SVC layered video is still missing in the literature. In particular, a question of practical importance is how this layered approach can benefit users with different reception capabilities. Furthermore, before this feedback-free approach can get acceptance in practice, a comparative study to quantify the performance degradation due to not using feedback is essential. Hence, to address these issues, the main focus in this paper will be on EW UEP+RLNC for live streaming of layered video with the aim to: 

\begin{itemize}
\item design optimal (in the sense we will describe below) feedback-free broadcast schemes over single-hop heterogeneous erasure channels,
\item compare the optimal feedback-free schemes with idealistic full-feedback schemes.
\end{itemize}
These are the main theoretical contributions of this paper.

We build upon the analysis in~\cite{Vukobratovic:2012:IEEE-TCOM:UEP} and start with the single-user case and then extend the study to the multi-user case. However, instead of the probabilistic approach for selecting coding windows for transmissions, we consider a deterministic approach, where the number of coded packets from each window is explicitly determined at the sender. In fact, the probabilistic approach in~\cite{Vukobratovic:2012:IEEE-TCOM:UEP} is used to cancel out the effect of erasure statistics, which is what we actually want to highlight and study for heterogeneous multi-user networks. Furthermore, with the deterministic approach, we are in fact reducing one level of uncertainty in our system design and implementations, thus our theoretical predictions are expected to be achieved with more robustness.


To design the optimal schemes, in contrast to studies that use PSNR as the design metric~\cite{Hulya:2009:IEEE-JSAC:VAO, Nguyen:2011:IEEE-TVT:JNC, Nazir:2012:ALS:h264}, we use and maximize a more objective, content-independent performance metric, which is defined based on the layer decoding probabilities and can reflect the expected percentage of decoded frames. This theoretical metric is advantageous over PSNR as its value does not depend on the actual content of video frames, but only on the number of packets of video frames, so it can be computed offline as look-up tables to save time during live streaming. 

The framework we propose for obtaining this theoretical performance metric is general and can be calculated for different video and system parameters, e.g., packet error rate (PER), number of packets per layer, number of layers and number of possible transmissions. In addition to studying the effect of different parameters, we propose an adaptive approach for selecting the optimum number of video layers for each group of picture (GOP, which is the building block of the encoded video streams). To this end, when performing fragmentation and aggregation of encoded video data to form packets and layers, we take into account the expected theoretical performance metric for different number of layers and choose the one that gives the best expected performance. This is the main practical contribution of this study.

To put our results into perspective and compare with full-feedback scheme as an upper-bound, we study an idealistic EW UEP+RLNC scheme, where perfect and immediate feedback about users' reception status is assumed to be available at the sender. We utilize the finite horizon MDP~\cite{Puterman:1994:MDP} to obtain the optimal performance of this idealistic scheme and then compare it with our proposed feedback-free scheme. We note that MDP has been utilized for streaming applications in different setups, e.g., for rate-distortion optimized streaming of packetized media in~\cite{Chou:2006:IEEE-TMM:RDO}, for dynamically optimized multi-user wireless video transmission in~\cite{Fu:2010:IEEE-JSAC} and for adaptive scheduling of stored scalable video in~\cite{Chen:2013:IEEE:MDP}; Here we adapt it to our problem of layered streaming using EW UEP+RLNC.

\ifCLASSOPTIONtwocolumn
\begin{figure*}
\centering
\includegraphics[width=5.8in]{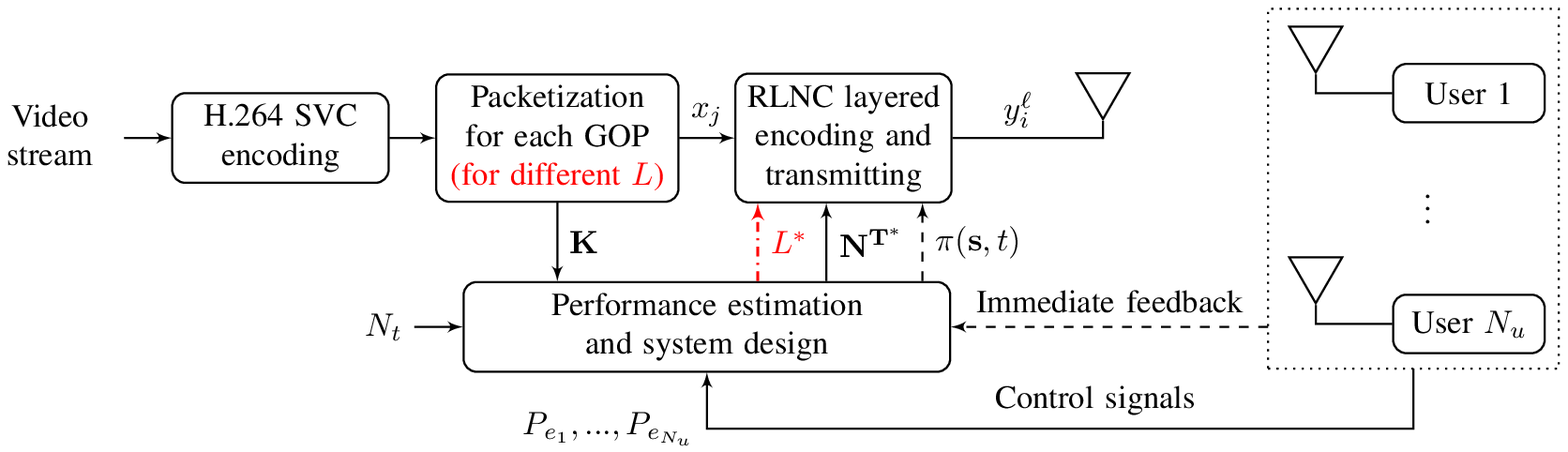}
\caption{System model showing different components and their relations. ${\bf N^{T^*}}$ is specific to feedback-free scheme, the immediate feedback and $\pi(s,t)$ shown in dashed line are specific to full-feedback scheme. Moreover, considering different number of layers $L$ and consequently the optimum one, $L^*$, is specific to opt-layer approaches. These are all discussed in corresponding future sections.} \label{Sys_model}
\end{figure*}
\fi

To assess the performance of the designed optimal schemes we use real H.264/SVC encoded video test streams and consider various systems parameters. We show the proposed feedback-free scheme performs very close to the idealistic scheme. Furthermore, the gain of using optimum number of video layers over using fixed number of video layers is clearly shown. Finally, to better study the effect of heterogeneity of users' erasure channels on the the defined theoretical performance metric, we incorporate the fairness of users' performances into the broadcast design and discuss the performance trade-offs.

The remainder of the paper is organized as follows. The system model, transmission schemes and the performance metrics are presented in Section II. In Section III, we formulate the decoding probabilities of different layers and define the theoretical performance metric for the single-user case. Then, we discuss the extension to the multi-user case in Section IV. Section V briefly describes the H.264/SVC video test streams and the PSNR calculation. The numerical results are provided in Section VI, and finally Section VII concludes the paper.


\section{System Model} \label{Section2}

The system model consists of a sender and a set of $N_u$ wireless users. The channels between the users and the sender are assumed to be independent and heterogeneous, i.e., they are not necessarily identical and have packet error rates (PERs) of $P_{e_i}, 1\leq i\leq N_u$. The sender is supposed to broadcast a live layered video stream to the users. We consider that the layered video data is chunked, where each chunk corresponds to a fixed number of frames that we refer to as a group of picture (GOP), to be compliant with video streaming standards.

Each GOP is composed of $M$ packets $x_1,..., x_M$, which are from $L$ layers. Layer $1$ is considered to be the most important layer and layer $L$ is the least important one. Packets of layer $\ell$ are useful only if all packets of lower layers are received (decoded) correctly. We consider layer $\ell$ to have $k_{\ell}$ packets, thus $\sum_{\ell=1}^Lk_{\ell}=M$. We use ${\bf K}=[k_1,k_2,...,k_L]$ to denote the number of packets from different layers in a GOP. For real video test streams, depending on the video content, $M$ and ${\bf K}$ can take different values for different GOPs. 

\ifCLASSOPTIONtwocolumn
\begin{table*} 
\renewcommand{\arraystretch}{1.35}
\caption{Comparing decoding outcome of the EW and NOW approaches for two erasure patterns examples.}
\label{Table_example}
\centering
\begin{tabular}{|c|cccccc||c|c|}
\hline
Transmissions&$y_1^1$&$y_2^1$&$y_3^1$&$y_1^2$&$y_2^2$&$y_3^2$& EW decoded?& NOW decoded?\\
\hline
 Erasure pattern 1&\cmark&\xmark &\xmark &\cmark &\cmark & \cmark & both layers & second layer, not useful\\
 Erasure pattern 2&\cmark&\cmark &\cmark &\xmark &\xmark & \cmark & first layer & first layer\\
\hline
\end{tabular}
\end{table*}
\fi

We apply our NC approach on the packets of each GOP as soon as they are all ready, which means neither merging of GOPs nor buffering of packets is considered. This is important in live streaming to minimize the delivery delay. For any GOP of interest, considering that the number of frames per GOP is $F$, video frame rate is $f$ frame per second (fps), the transmission rate is $r$ bit per second (bps) and the selected packet length is $n$ bits, the possible number of packet transmissions for one GOP is fixed and limited. We denote this number by $N_t$ and it can be easily inferred that $N_t=\tfrac{Fr}{nf}$.  

A schematic model of the considered system is depicted in Fig.~\ref{Sys_model}. Different system blocks are discussed in corresponding sections: \emph{RLNC layered encoding and transmitting} in Sections~\ref{RLNC} and~\ref{TX_scheme}, \emph{Performance estimation and system design} in Sections~\ref{Formulation_single} and~\ref{Formulation_multi}, and \emph{H.264/SVC encoding and Packetization} in Section~\ref{SVC_PSNR}.


\ifCLASSOPTIONonecolumn
\begin{figure*}
\centering
\includegraphics[width=5.8in]{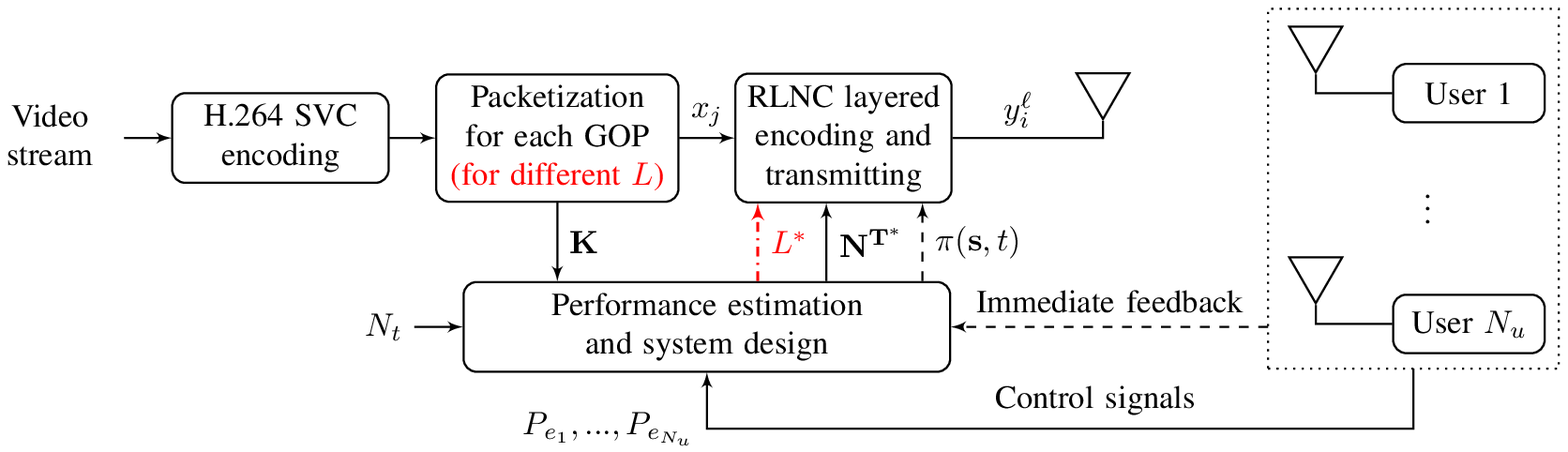}
\caption{System model showing different components and their relations. ${\bf N^{T^*}}$ is specific to feedback-free scheme, the immediate feedback and $\pi(s,t)$ shown in dashed line are specific to full-feedback scheme. Moreover, considering different number of layers $L$ and consequently the optimum one, $L^*$, is specific to opt-layer approaches. These are all discussed in corresponding future sections.} \label{Sys_model}
\end{figure*}
\fi

\subsection{Random Linear Network Coding (RLNC)}
\label{RLNC}

\begin{figure}
\centering
\includegraphics[width=2.4in]{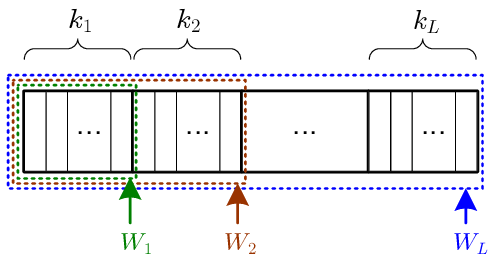}
\caption{An $L$-layer GOP with $k_\ell$ packets in the $\ell$-th layer. Examples of the expanding windows are shown.} \label{Fig01}
\end{figure}

Similar to~\cite{Vukobratovic:2012:IEEE-TCOM:UEP}, we utilize RLNC~\cite{Ho:IEEE-IT:2006:RLN} in this study and use coding over expanding windows. In this approach, $L$ coding windows are considered, where $\ell$-th window, denoted by $W_\ell$, contains all packets from layers $1$ to $\ell$ as shown in Fig.~\ref{Fig01}. Then, based on the transmission policy, network coded packets from different windows are generated and transmitted. This approach is often considered in contrast to coding over non-overlapping windows (NOW)~\cite{Vukobratovic:2012:IEEE-TCOM:UEP}, where coding window $W_\ell$ contains only packets of layer $\ell$. In fact, coding using NOW considers only intra-layer NC, but coding using EW allows for inter-layer NC in addition to intra-layer NC. Consequently, in the EW approach, packets of more important layers are protected better and are more likely to get decoded compared to the NOW approach, as shown in the upcoming Example~\ref{Example1}.

The theory behind RLNC encoding and decoding has been studied comprehensively in the literature during the past decade, with the effect of field size $q$ also discussed quite in detail (e.g.,~\cite{Lucani:2012:IEEE-IT:NCD,Trullols:2011:IEEE-Comm}). Hence, we do not elaborate on these issues here. We only assume that the encodings are over large enough field sizes, which leads us to the following remarks: 

\begin{remark} \label{remark1}
The coded packets generated from packets of a coding window are all linearly independent (with high probability, which we approximate to be one).
\end{remark}
\begin{remark} \label{remark2}
Considering the EW approach, in order to decode all the packets of a coding window $W_{\ell_1}$, $\sum_{\ell=1}^{\ell_1}k_{\ell}$ linearly independent coded packets from $W_{\ell_1}$ window are required. With the assumption of large enough field sizes, coded packets from smaller windows, i.e., $W_{\ell_2}$ with $1\leq \ell_2<\ell_1$ can also be used in decoding, but with a limitation that for every $\ell_2$, the total number of coded packets from windows $W_1$ to $W_{\ell_2}$ does not exceed $\sum_{\ell=1}^{\ell_2}k_{\ell}$.
\end{remark}

These remarks are used to obtain the decoding probabilities for our approach in the next section.
Note that the effect of field size $q$ in decoding probabilities can be later incorporated in our study, as has been discussed in~\cite{Vukobratovic:2012:IEEE-TCOM:UEP, Esmaeilzadeh:2014:IEEE-TCOM:JOP}.

To make the remarks more clear, let us consider the following simple example that also provides comparison between EW and NOW approaches.

\begin{exmp} \label{Example1}
In this example, we consider $L=2$ layers and assume that $k_1=k_2=2$, i.e., packets $x_1$ and $x_2$ are from the first layer and packets $x_3$ and $x_4$ are from the second layer. We use $y_i^\ell$ to denote the $i$-th coded packet from the $\ell$-th window, which is obtained as $y_i^\ell=\sum_{x_j\in W_\ell} a_{i,j}x_j$. Here, $W_1$ contains packets $x_1$ and $x_2$ for both the EW and NOW approaches, $W_2$ contains all the four packets in the EW approach, but only packets $x_3$ and $x_4$ in the NOW approach, and $a_{i,j}$'s are the randomly chosen coefficients from $\mathbb{F}_q$. Considering that $N_t=6$ transmissions are possible, Table~\ref{Table_example} provides two different erasure patterns examples and compares the decoding outcomes for the EW and NOW approaches when three coded packets from each of the windows are transmitted. 

Considering the first erasure pattern, it can be seen that EW can decode packets of both layers and thus outperforms the NOW approach that only decodes packets of the second layer, which are useless without packets of the first layer. Moreover, considering the two different erasure patterns for the EW approach, although $4$ coded packets are received in total in both cases, only in the first case both layers are decodable, which is directly concluded from Remarks~\ref{remark1} and~\ref{remark2}.  
\qed
\end{exmp}

\ifCLASSOPTIONonecolumn
\begin{table*} 
\renewcommand{\arraystretch}{1.25}
\caption{Comparing decoding outcome of the EW and NOW approaches for two erasure patterns examples.}
\label{Table_example}
\centering
\begin{tabular}{|c|cccccc||c|c|}
\hline
Transmissions&$y_1^1$&$y_2^1$&$y_3^1$&$y_1^2$&$y_2^2$&$y_3^2$& EW decoded?& NOW decoded?\\
\hline
 Erasure pattern 1&\cmark&\xmark &\xmark &\cmark &\cmark & \cmark & both layers & second layer, not useful\\
 Erasure pattern 2&\cmark&\cmark &\cmark &\xmark &\xmark & \cmark & first layer & first layer\\
\hline
\end{tabular}
\end{table*}
\fi

\subsection{Transmission Schemes}
\label{TX_scheme}

In this paper, we consider two transmission schemes both using the EW approach, a feedback-free RLNC-based transmission scheme, and also an idealistic full-feedback RLNC-based transmission scheme. The purpose is to investigate how close the proposed feedback-free scheme can perform compared to the idealistic upper-bound. Here, the mentioned notion of feedback is for the RLNC methods and not for the whole system. In other words, as shown in Fig.~\ref{Sys_model}, even for the feedback-free scheme, there exist backward links from users to the sender to communicate control signals and statistical information (e.g., to estimate $P_{e_i}$) infrequently as long as the total bandwidth requirements of backward links are satisfied.

As mentioned previously, based on the system parameters, we assume the sender can transmit at most $N_t$ NC packets for each GOP. Here we explain how these transmissions are carried out for each of these schemes and will formulate their performance in the next section. 

\subsubsection{Feedback-free Scheme} 

In this scheme, for each GOP, the sender decides in advance on the number of coded packets from each coding window that should be transmitted, and then sends them one after another, without waiting for any feedback. Assuming that $n_\ell^t$ packets are generated (and thus transmitted) from the packets in the $\ell$-th window, then $\sum_{\ell=1}^Ln_\ell^t=N_t$ and we call ${\bf N^T}=[n_1^t,n_2^t,...,n_L^t]$ a feedback-free transmission policy. The decision on the optimum policy is made based on an aggregate function of users' performance by taking into account $N_t$ and channels characteristics. This will be discussed in Section~\ref{Formulation_multi}.
 
\subsubsection{Full-feedback Scheme}

In this idealistic scheme, it is assumed that the sender, before every transmission, knows exactly how many packets of different windows each user has received so far. Hence, based on this information, channels characteristics, and the remaining number of transmissions from total $N_t$, the sender decides on the next transmission to optimize an aggregate function of users' performance. Since we have considered $L$ windows for generation of coded packets, the sender should in fact decide on the coding window from which the next packet for transmission is generated.

\subsection{Performance Metrics}

In this paper, we will consider two types of performance metrics. The main metric that we use for our systems designs, is the weighted sum of the layer decoding probabilities, which is theoretical and content-independent, i.e., does not depend on the actual video content, but on the number of video packets. Based on the selection of the weights, this metric can reveal, for example, the expected percentage of decoded frames or the expected throughput. The formulation of this metric is presented in Sections~\ref{Formulation_single} and~\ref{Formulation_multi}. In the results section, in addition to obtaining the average values for the theoretical performance metric, we also calculate the average PSNR as it is a widely used metric for video quality. PSNR calculations are described in Section~\ref{SVC_PSNR}.

\section{Formulation of Theoretical Performance Metrics -- Single-user case} \label{Formulation_single}

In this section, we present the formulation of the theoretical performance metric of the feedback-free and full-feedback schemes for the single-user case, and then will discuss their extension to the multi-user case in the next section. These two sections in fact build the framework for \emph{Performance estimation and system design} block illustrated in Fig.~\ref{Sys_model}.

\subsection{Feedback-free Scheme} \label{feedback_free_1user}

Under the assumptions made in Section~\ref{Section2}, we start with formulating the probability that a user with PER of $P_e$ can decode the packets of layer $\ell$ (and of course all the packets of lower layers). We denote this probability by $P_\ell({\bf K},{\bf N^T})$. Then, we define the expected theoretical performance metric as the weighted sum of these probabilities for all $1\leq \ell \leq L$.

To obtain these probabilities, we assume that out of the $n_\ell^t$ coded packets of the $\ell$-th layer, $n_\ell^r$ packets are received, where $0\leq n_\ell^r\leq n_\ell^t$, thus we denote by ${\bf N^R}=[n_1^r,n_2^r,...,n_L^r]$ the number of received packets from different layers. Then, considering all possible ${\bf N^R}$, $P_\ell({\bf K},{\bf N^T})$ can be written as
\begin{align} \label{feedback-free_P_ell}
P_\ell({\bf K},{\bf N^T}) = \sum_{\text{all possible }{\bf N^R}}P({\bf N^R}|{\bf N^T})I(L_{max}({\bf K},{\bf N^R})=\ell)
\end{align}
where
\begin{align}
P({\bf N^R}|{\bf N^T}) = \prod_{\ell=1}^L{{n_\ell^t}\choose{n_\ell^r}}(1-P_e)^{n_\ell^r} P_e^{n_\ell^t-n_\ell^r}
\end{align}
Here, $I(\cdot)$ is an indicator function with output $1$ if its argument, which is a logical expression, is true. 

The function $L_{max}({\bf K},{\bf N^R})$ calculates the highest decodable layer based on Remarks~\ref{remark1} and~\ref{remark2}. The value for this function can be calculated as follows:
\begin{align} \label{feedback-free_LMax}
L_{max}({\bf K},{\bf N^R}) = \max\{D(1), D(2), ..., D(L)\}
\end{align}
where $D(\ell)$ is equal to $\ell$ when packets of window $\ell$ are all decodable and zero otherwise. Hence, $D(1)=I(n_1^r\geq k_1)$ and $D(\ell)$ for $2\leq \ell \leq L$ are calculated using the following recursive formulas:
\begin{align}
b(\ell) = \max\{D(1),... ,D(\ell-1)\}
\end{align}
\begin{align} \label{Dec_condition}
D(\ell) = \ell \times I\left(\sum_{i=b(\ell)+1}^\ell n_i^r \geq \sum_{i=b(\ell)+1}^\ell k_i\right)
\end{align}
In fact in every iteration, $b(\ell)$ holds the index of the largest decodable window from previous iterations, and then the decoding condition in~\eqref{Dec_condition} is tested for the remaining undecoded packets. Some sample outputs of the $L_{max}({\bf K},{\bf N^R})$ function are presented in the following example and the calculation steps for one of the samples are given in Table~\ref{Table_example2}.

\begin{exmp}
Considering ${\bf K}=[5,1,2,3]$, the outputs of $L_{max}({\bf K},{\bf N^R})$ function for ${\bf N^R}=[4,1,2,3]$, ${\bf N^R}=[5,0,2,3]$, ${\bf N^R}=[4,3,1,3]$, ${\bf N^R}=[0,4,4,2]$ and ${\bf N^R}=[3,0,0,8]$ are $0$, $1$, $2$, $3$ and $4$, respectively.
\begin{table} [!h]
\renewcommand{\arraystretch}{1.35}
\caption{Steps for calculation of $L_{max}({\bf K},{\bf N^R})$ using ${\bf K}=[5,1,2,3]$ and ${\bf N^R}=[4,3,1,3]$.}
\label{Table_example2}
\centering
\begin{tabular}{|c|cccc|}
\hline
$\ell$&$b(\ell)$& decoding condition& condition met? &$D(\ell)$
\\
\hline
1&-&$4\geq 5$&\xmark&0 \\
2&0&$4+3\geq 5+1$&\cmark&2 \\
3&2&$1\geq2$&\xmark&0 \\
4&2&$1+3\geq 2+3$&\xmark&0 \\
\hline
\end{tabular}
\end{table}
\vspace{-5mm}
\end{exmp}
     \qed  \\
 
Having calculated the layer decoding probabilities, the theoretical performance metric can be defined as follows:
\begin{align} 
\eta = \sum_{\ell=1}^Lc_\ell P_\ell({\bf K},{\bf N^T}) \label{feedbackfree_eta}
\end{align}
where $c_\ell$ reflects the cumulative importance of layers $1$ to $\ell$. For instance, considering the temporal scalability in SVC of H.264, for a $2$-layer case, if the number of frames per layer are equal, with $c_1=0.5$ and $c_2=1$, $\eta$ will give the expected percentage of decoded frames. The explanations in Section~\ref{SVC_PSNR} will better clarify this. As another example, if we can consider $c_\ell$ to be the ratio of the number of packets obtained by decoding layer $\ell$ to the total number of packets as
\begin{align} \label{coef_throughput}
c_\ell=\frac{\sum_{j=1}^\ell k_j}{\sum_{j=1}^{L}k_j}\text{,}
\end{align}
then $\eta$ will give the expected throughput. We emphasize that the metric defined in~\eqref{feedbackfree_eta} is a general metric and caters for any other weighting as required by the application.

With the theoretical formulation for $\eta$ obtained in \eqref{feedbackfree_eta}, the optimal feedback-free policy for the single-user case can be acquired by 
\begin{align} \label{single_opt_feedbackfree}
{{\bf N^T}}^* = \arg \max_{n_1^t,...,n_L^t}\{\eta\}\text{,~~subject to}~\sum_{\ell=1}^Ln_\ell^t=N_t 
\end{align}
which is solved in Section~\ref{RESULTS} by exhaustively searching through all possible combinations of $[{n_1^t,...,n_L^t}]$ that are at most $(N_t)^L$ cases. Assuming the time complexity of calculating $\eta$ (using~\eqref{feedback-free_P_ell}-\eqref{feedbackfree_eta}) for different ${\bf N^T}$ is upper-bounded by $\Gamma_\eta(N_t,L)$, the optimization in~\eqref{single_opt_feedbackfree} has overall time complexity smaller than $O(\Gamma_\eta(N_t,L).(N_t)^L)$.
\subsection{Full-feedback Scheme} \label{fullFeedback_single}

In this section, the objective is to obtain the formulation of a similar theoretic performance metric as in \eqref{feedbackfree_eta} for the full-feedback scheme. To this end, we utilize the \emph{finite horizon Markov decision process (MDP)}~\cite{Puterman:1994:MDP}.

In the full-feedback scheme, as mentioned previously, before every transmission the sender should decide on what to transmit next. The decision is made to optimize the performance metric by considering the immediate information about the reception status of the user, the channel status, as well as the remaining number of transmissions. Here, since the number of transmissions $N_t$ is limited, this can be best modeled by the finite horizon MDP, with $N_t$ horizons (stages). A summary of finite horizon MDP is presented in Appendix~\ref{Appendix_MDP}. Here, we will explain how the different components of MDP should be assigned in our considered scenario.

\subsubsection{States}
For the states, we consider an $L$-tuple ${\bf s}=(d_1,d_2,...,d_L)$, where each $0\leq d_\ell\leq k_\ell$ element shows the remaining required independent packets to decode layer $\ell$. Thus, it is obvious that at the start of the transmission, the state is ${\bf s^0}=(k_1,k_2,...,k_L)$ and the state space $\mathcal{S_{\text{single}}}$ has a size of  $|\mathcal{S_{\text{single}}}|=\prod_{\ell=1}^L(k_\ell+1)$.

\subsubsection{Actions}
\begin{figure}
\centering
\ifCLASSOPTIONonecolumn
\includegraphics[width=3.3in]{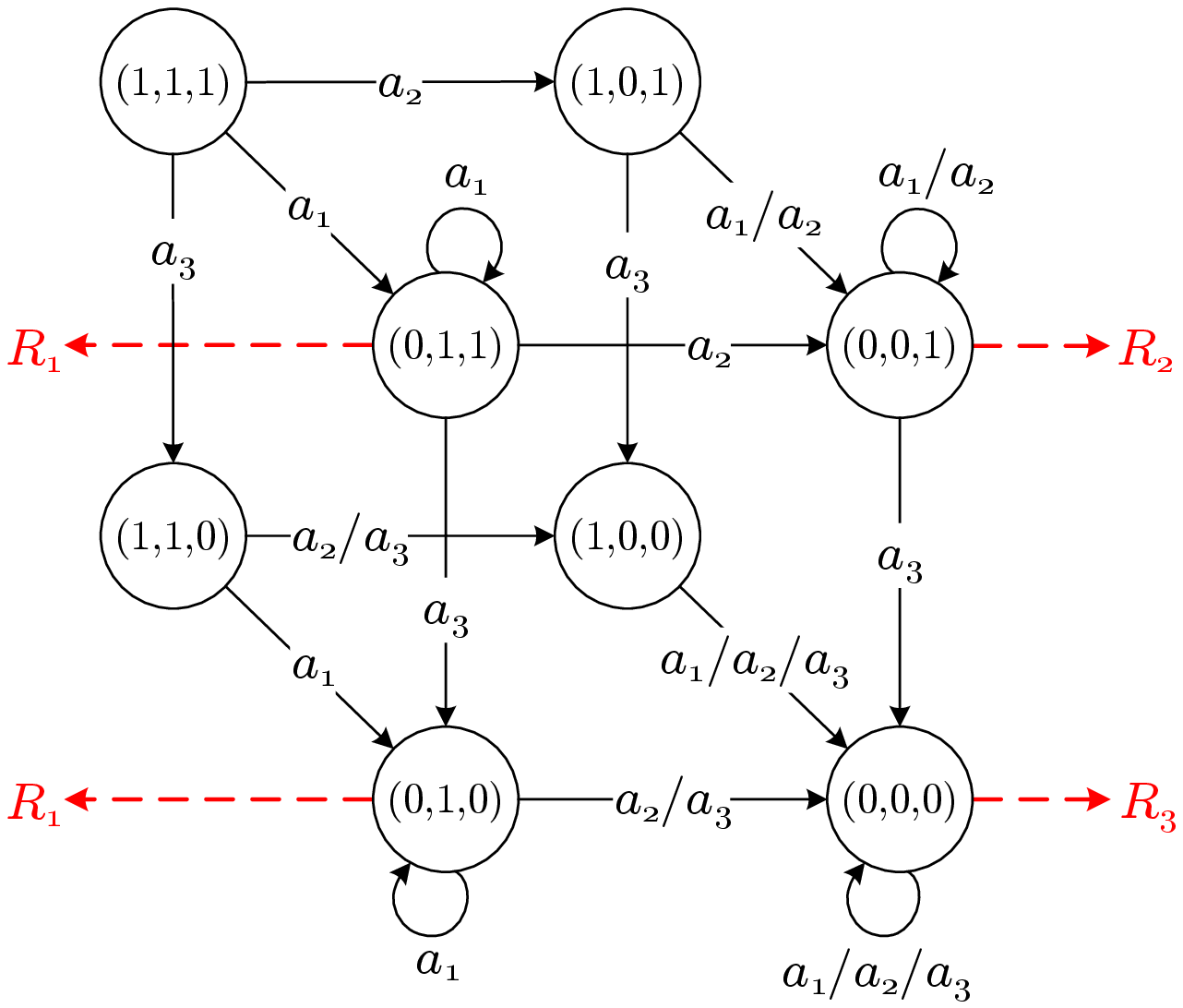}
\else
\includegraphics[width=3in]{Action_states_new.eps}
\fi
\caption{An example of states, actions and terminal rewards. Terminal rewards for other states are zero. } \label{Fig1}
\end{figure}

For the actions, we consider $L$ different actions $\mathcal{A} = \{a_1,...,a_L\}$, where in action $a_\ell$, a coded packet generated from packets of the coding window $W_\ell$ is transmitted. Under action $a_\ell$, if the coded packet is received correctly (with probability $1-P_e$), then a change in state occurs if at least one of the $d_1$ to $d_\ell$ is nonzero. Otherwise, if all of the $d_1$ to $d_\ell$ are zero, state will not change. In fact, the latter shows a case where all the packets of the first $\ell$ layers are successfully decoded, so a coded packet from the $\ell$-th layer does not give any extra information. It is apparent that in the case of not receiving the coded packet (with probability $P_e$), the state remains unchanged. An example of this is shown in Fig.~\ref{Fig1}, assuming the coded packet is received, $L=3$ and starting state ${\bf s^0}=(1,1,1)$.

\subsubsection{State transition probabilities}
Now, considering the state transitions caused by actions, we first define the state transition function with the assumption of successful reception of coded packets and then obtain the state transition probabilities. We assume the current state is ${\bf s}=(d_1,d_2,...,d_L)$ and want to obtain next state ${\bf s^n}=(d_1^n,d_2^n,...,d_L^n)$ following an action $a_\ell$. Function $F({\bf s},a_\ell)$ calculate ${\bf s^n}$ as follows:
\ifCLASSOPTIONonecolumn
\begin{align} \label{transition_func_full}
{\bf s^n} =F({\bf s},a_\ell)=\left\{
\begin{array}{lll}
(d_1,...,d_\ell-1,...,d_L)&;& d_\ell\neq 0 \\
(d_1,...,d_{i^*}-1,...,d_L)&; & d_\ell=0~\&~\exists~i \in \{1,...,\ell-1\}: d_i \neq0 \\
&& i^*=\max\{i\}\\
(d_1,...,d_L)&;& \forall ~i \in \{1,...,\ell\}: d_i=0
\end{array}\right.
\end{align}
\else
\begin{align} \label{transition_func_full}
&{\bf s^n} =F({\bf s},a_\ell)= \nonumber \\
&\left\{
\begin{array}{lll}
(d_1,...,d_\ell-1,...,d_L)&;& d_\ell\neq 0 \\
(d_1,...,d_{i^*}-1,...,d_L)&; & d_\ell=0~\& \\
&& \exists~i \in \{1,...,\ell-1\}: d_i \neq0 \\
&& i^*=\max\{i\}\\
(d_1,...,d_L)&;& \forall ~i \in \{1,...,\ell\}: d_i=0
\end{array}\right.
\end{align}
\fi

To clarify how this function works, we use the example in Fig.~\ref{Fig1} and consider state $(1,1,0)$. Following actions $a_1$ and $a_2$, transitions to states $(0,1,0)$ and $(1,0,0)$, respectively, are straightforward as both $d_1$ and $d_2$ are nonzero. For action $a_3$, since $d_3$ is zero, from $d_2$ and $d_1$, $d_2$ is reduced by one for the next state, as it has the largest index among the nonzero $d_\ell$. This is to assure that enough independent packets are received before decoding of layers and is better understood in conjunction with the definition of terminal reward functions in Section~\ref{RTRF}.   

Using $F({\bf s},a_\ell)$, the state transition probabilities $P_{\text{single}}({\bf \acute{s}}|{\bf s},a_\ell)$ for any ${\bf s},{\bf \acute{s}}\in \mathcal{S_{\text{single}}}$ and $a_\ell\in \mathcal{A}$ can be obtained by
\begin{align}
P_{\text{single}}({\bf \acute{s}}|{\bf s},a_\ell) = (1-P_e)I({\bf \acute{s}}=F({\bf s},a_\ell)) + P_e I({\bf \acute{s}}={\bf s})
\end{align}

\subsubsection{Reward and terminal reward functions} \label{RTRF}

We consider the reward function $R({\bf s},a)$ to be zero for all ${\bf s}\in \mathcal{S_{\text{single}}}$ and $a\in \mathcal{A}$ and use only the terminal reward $G_{\text{single}}({\bf s})$ to capture the effect of layer decoding on the theoretical performance metric. This approach is valid because the state in which the user is when the transmission has finished is important in calculation of the performance metric in this scheme, but not how or when it has reached that state. Hence, using the same idea shown in Fig.~\ref{Fig1} for terminal rewards, the terminal reward can be obtained, in general, as follows:
\begin{align}
G_{\text{single}}({\bf s}) = \left\{
\begin{array}{lll}
R_{\ell^*}&;& \exists~\ell \in \{1,...,L\}: \sum_{i=1}^\ell d_\ell=0 \\
&& \ell^*=\max\{\ell\}\\
0&;& \text{otherwise}
\end{array}\right.
\end{align}

Here, reward of $R_1$ means that only packets of the first layer are decoded and reward of $R_\ell$ means that packets of layer $\ell$ and all the lowers layer are decoded. 

Back to the example in Fig.~\ref{Fig1}, it can be observed that if we start from state $(1,1,1)$, in the case of one successful action $a_1$ and one successful action $a_3$ (regardless of the order), we end up in state $(0,1,0)$ and only first layer is decodable. Hence, reward of $R_1$ is given. In another case, with only two successful $a_3$ actions, the final state will be $(1,0,0)$ and even the first layer cannot be decoded. Thus reward of zero is considered. 

Having discussed all the elements of finite horizon MDP, the optimal theoretical performance metric will be the value function\footnote{$V_{\pi}^{t}(s)$, this is defined in Appendix~\ref{Appendix_MDP}.} at stage $1$ (i.e., $N_t$ transmissions to go) for state ${\bf s^0}=(k_1,k_2,...,k_L)$ defined as
\begin{align}
\eta = V_{\pi}^{N_t}({\bf s^0}) \label{PM2}
\end{align}

Note that, in contrast to the formulation for the feedback-free scheme~\eqref{single_opt_feedbackfree} that required a single optimization to obtain the optimal policy, the proposed solution for the full-feedback scheme, which is based on finite horizon MDP, requires an optimization at every stage, hence the $\eta$ given in~\eqref{PM2} is optimal and the optimal policy is in fact given by the obtained optimal actions for different states and stages, i.e., $\pi({\bf s},t)$ for ${\bf s}\in \mathcal{S_{\text{single}}}$ and $1\leq t\leq T$. As discussed in Appendix~\ref{Appendix_MDP}, this method has time complexity of $O(N_tL|\mathcal{S_{\text{single}}}|^2)$. In order for the metric in~\eqref{PM2} to measure values with similar meaning as in~\eqref{feedbackfree_eta}, it is required that $R_\ell = c_\ell$.

\section{Formulation of Theoretical Performance Metrics -- Multi-user case} \label{Formulation_multi}

In this section, the objective is to discuss how the formulations in the previous section can be used and extended to the multi-user case in order to design RLNC-based video streaming. As stated in the system model, $N_u$ wireless users with independent and heterogeneous erasure channels are considered and the purpose is to optimize an aggregation of their performance metrics. 

\subsection{Feedback-free Scheme}
For this scheme, the performance metric of user $i$ with $P_{e_i}, 1\leq i\leq N_u$, which is  denoted by $\eta_i$, can be obtained independently by using~\eqref{feedbackfree_eta} for any ${\bf K}$ and ${\bf N^T}$. Then the aggregate performance metric is defined as:
\begin{align}
\eta_{tot} = H(\eta_1,...,\eta_{N_u}) \label{Free_Multi1}
\end{align}
where $H(\cdot)$ is the considered aggregate function. Hence, the optimal policy will be obtained as
\begin{align} 
{\bf N^T}^* = \arg \max_{n_1^t,...,n_L^t}\{\eta_{tot}\}\text{,~~subject to}~\sum_{\ell=1}^Ln_\ell^t=N_t  \label{Free_Multi2}
\end{align}

We note that various functions can be considered for $H(\cdot)$, such as mean, geometric mean, or even functions that consider the performances of a subset of users. The decision about this is made based on the network configuration and the requirements of applications.

\subsection{Full-Feedback Scheme}

The extension of the single-user's formulation to multi-user case for the full-feedback scheme is more complicated compared to the feedback-free scheme discussed in the previous subsection. This is due to the fact that in the full-feedback scheme, decisions about the coded packets to be transmitted are made based on the reception status of all or a subset of users before every transmission. Hence, the performance of each user, in addition to its own reception, is dependent upon the reception of other users. This makes the finite horizon MDP problem more complicated.

To obtain the input components of the finite horizon MDP, we consider the same function $H(\cdot)$ to calculate the aggregate performance metric. Considering that the performances of a subset of $n_u$ users (out of $N_u$) are taken into account in the sender's decisions ($1\leq n_u \leq N_u$), it can be easily inferred that the multi-user state space $\mathcal{S}_{\text{multi}}$ has a size of $|\mathcal{S}_{\text{multi}}|=|\mathcal{S_{\text{single}}}|^{n_u}$. Then, a state ${\bf s}\in\mathcal{S}_{\text{multi}}$ is defined with an $n_u$-tuple $({\bf s_{1}},...,{\bf s_{n_u}})$, where each ${\bf s_{j}}\in \mathcal{S_{\text{single}}}$ is itself an $L$-tuple, as defined in Section~\ref{fullFeedback_single}, showing the state for user $j$, $1\leq j\leq n_u$. 

Since the actions are similar to the single-user case, the next component to be computed is the transition probability function. For any ${\bf s},{\bf \acute{s}}\in \mathcal{S_{\text{multi}}}$, the transition probability function $P_{\text{multi}}({\bf \acute{s}}|{\bf s},a)$ can be calculated as
\begin{align}
P_{\text{multi}}({\bf \acute{s}}|{\bf s},a)=\prod_{j=1}^{n_u}P_{\text{single}}({\bf \acute{s_j}}|{\bf s_j},a)
\end{align}
where ${\bf s_j},{\bf \acute{s_j}}\in \mathcal{S_{\text{single}}}$. In fact, the state transitions caused by action $a$ are independent for different users, thus the multiplication of the single-user transition probability functions $P_{\text{single}}({\bf \acute{s_j}}|{\bf s_j},a)$ gives the multi-user transition probability function.

For the reward functions, we again assume that $R({\bf s},a)$ has zero value for all the actions and states. Then, in order to properly model the reward in every state ${\bf s}\in \mathcal{S_{\text{multi}}}$, the terminal reward $G_{\text{multi}}({\bf s})$ should be defined as follows:
\begin{align}
G_{\text{multi}}({\bf s}) = H(G_{\text{single}}({\bf s_1}),...,G_{\text{single}}({\bf s_{n_u}}))
\end{align}

Having defined all the components for the multi-user finite horizon MDP, the optimal theoretical performance metric is the value function at stage 1 for state ${\bf s_{\text{multi}}^0}=({\bf s_1^0},...,{\bf s_{n_u}^0})$, where ${\bf s_j^0}=(k_1,k_2,...,k_L)$ for every user $1\leq j \leq n_u$. Hence, the aggregate performance metric is given by the value function at stage 1 for state ${\bf s_{\text{multi}}^0}$, as follows
\begin{align}  
\eta_{tot}=V_\pi^{N_t}({\bf s_{\text{multi}}^0}) \label{MDP_Multi}
\end{align}
which requires calculation of $V_\pi^t({\bf s})$ and $\pi({\bf s},t)$ for every ${\bf s}\in \mathcal{S_{\text{multi}}}$ and $1\leq t \leq N_t$.

\subsection{On the Computational Complexities of Multi-user Schemes}

In this subsection, we briefly discuss the computational complexities of the feedback-free and full-feedback schemes  when $n_u$ users (out of $N_u$) are considered for multi-user system design.

Regarding the feedback-free scheme, since $\eta_i$ can be calculated independently for different users, the complexity increases linearly with $n_u$. Hence, solving the optimization in~\eqref{Free_Multi2} exhaustively has time complexity smaller than $O(n_u\Gamma_\eta(N_t,L).(N_t)^L)$.

For the full-feedback scheme, as the size of the state space increases exponentially, the computational complexity of obtaining the optimum policies (actions) also grows exponentially, i.e., $O(N_tL|\mathcal{S_{\text{single}}}|^{2n_u})$.

While the exponential complexity is not desirable, we emphasize that the proposed full-feedback scheme is an idealistic scheme used as a benchmark. Furthermore, although we will select $n_u$ to be equal to the total number of users ($N_u$) in our simulations, it is possible to judiciously design the system based on a subset of users, as in~\cite{Lucani:2012:IEEE-IT:NCD}, to keep the complexity reasonable. Moreover, we emphasize that many of the optimization steps that require demanding computations do not need to be calculated online for every GOP. Instead, they can be tabulated (as look-up tables, LUTs) for expected more common system parameters offline, or LUTs can even be gradually filled as the system is trained.

\section{SVC Video Streams and PSNR Calculations} \label{SVC_PSNR}

This section provides details about the H.264/SVC video test streams used in this paper and also briefly explains the PSNR calculations.

\begin{figure}
\centering
\ifCLASSOPTIONonecolumn
\includegraphics[width=3.5in]{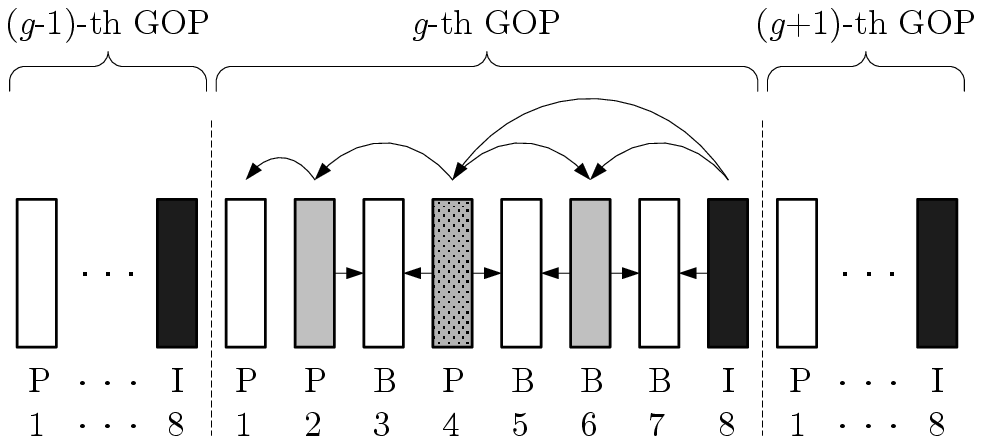}
\else
\includegraphics[width=3.3in]{GOP.eps}
\fi
\caption{A closed GOP with 8 frames, constituted by I, P and B frames.} \label{Fig_GOP}
\end{figure}

We consider three standard video streams: \emph{Foreman}, \emph{Crew} and \emph{Soccer}~\cite{test_video}. These streams are all in common intermediate format (CIF, i.e., $352\times288$) and all have $300$ frames with $30$ fps. We encode them using the JSVM 9.19.14 version of H.264/SVC codec~\cite{JSVM-SVC, Schwarz:2007:IEEE:SCV} by considering GOP size of $F=8$ and benefiting only from the temporal scalability of H.264\footnote{The proposed framework for calculating the expected theoretical performance metric is general and can be easily applied to other scalability types (e.g., spatial and quality). Furthermore, as the framework treats each GOP separately, GOPs with variable number of frames can also be considered.}. This results in $38$ GOPs for each stream, each GOP composed of a sequence of I, P and B frames that can be considered into $4$ layers at most. This is shown in Fig.~\ref{Fig_GOP} by using gray shades, where the darker the shade, the more important the frame(s).

We note that the I frames are coded without reference to any frame except themselves, the P frames are predicted/coded with reference to one other frame, either an I or a P frame, and the B frames are bi-directionally predicted/coded, i.e., with reference to two other frames, e.g., one I and one P frame, two P frames or one I and one B frame. Hence, we can say the GOP shown in Fig.~\ref{Fig_GOP} is a closed GOP, which means the decoding of the frames inside the GOP is independent of frames outside the GOP.

Having obtained the encoded frames shown in Fig.~\ref{Fig_GOP}, the information bits should be assigned to the layers and packets. To this end, we consider the maximum transmission unit (MTU) of $n=1500$ bytes as the packet length, which is the largest allowed packet over Ethernet, and do aggregation and fragmentation of information bits of different frames to form packets. We dedicate $100$ bytes to all the header information, thus $n_{eff}=1400$ bytes are used for video data. Considering that the encoded frame $i, 1\leq i\leq 8$, is composed of $m_i$ bytes, the easiest way to assign video information to layers and packets would be the 1-layer case with $k_1=\lceil \tfrac{\sum_{i=1}^8m_i}{n_{eff}} \rceil$ packets. For the 2-layer case, we choose $k_1=\lceil \tfrac{m_2+m_4+m_6+m_8}{n_{eff}} \rceil$ and $k_2=\lceil \tfrac{m_1+m_3+m_5+m_7}{n_{eff}} \rceil$. For the 3-layer case, packets per layer can be considered as $k_1=\lceil \tfrac{m_4+m_8}{n_{eff}} \rceil$, $k_2=\lceil \tfrac{m_2+m_6}{n_{eff}} \rceil$  and $k_3=\lceil \tfrac{m_1+m_3+m_5+m_7}{n_{eff}} \rceil$ and finally for the 4-layer case, we consider $k_1=\lceil \tfrac{m_8}{n_{eff}} \rceil$, $k_2=\lceil \tfrac{m_4}{n_{eff}} \rceil$, $k_3=\lceil \tfrac{m_2+m_6}{n_{eff}} \rceil$  and $k_4=\lceil \tfrac{m_1+m_3+m_5+m_7}{n_{eff}} \rceil$.  

\begin{remark} \label{remark3}
It should be noted that due to using the ceiling function $\lceil\cdot\rceil$, for any GOP of interest, the sum of calculated $k_{\ell_1}$ for an $L_1$-layer case ($\ell_1\leq L_1$) can be smaller than the sum of calculated $k_{\ell_2}$ for an $L_2$-layer case ($\ell_2\leq L_2$) when $L_1<L_2$. In other words, the total number of packets $M$ is affected by the number of layers $L$. This will have an effect on the performance metrics too. Therefore, in addition to the approaches that use fixed number of layers, which we refer to as `$L$-layer' approaches, we also study an approach that adaptively selects the optimum number of layers based on $N_t$ and ${\bf K}$ for each GOP such that the highest theoretical performance is achieved for that GOP. This is referred to as `opt-layer' approach and its performance is later discussed in Section~\ref{real_single}.
\end{remark}

\begin{remark} \label{remark4}
The defined 1-layer case only offers nominal temporal resolution of $30$ fps for each GOP, i.e., either all 8 frames of a GOP are decoded or none are decoded, whereas the other cases provide more temporal resolutions, e.g., for the 4-layer case, based on the decoded number of layers, nominal temporal resolution of $3.75$, $7.5$, $15$ and $30$ fps are possible for each GOP, which correspond to 1, 2, 4 or 8 decoded frames out of 8. Hence, back to performance metric calculation in Section~\ref{feedback_free_1user}, to obtain the expected percentage of decoded frames, $c_\ell$ should be selected as $c_1=1$ for the $1$-layer case and $c_1=1/8$, $c_2=1/4$, $c_3=1/2$ and $c_4=1$ for the $4$-layer case, which can be written in the general form of $c_\ell = 2^{\ell-L}$. 
\end{remark}

To utilize PSNR as a performance metric, we use the luminance (Y) component of video sequences and obtain $\sigma_{i_1,i_2}$, which is the Y-PSNR if the uncompressed $i_1$-th frame is replaced by the compressed (i.e., encoded) $i_2$-th frames (for $1\leq i_1,i_2 \leq 300$). Considering ${\bf Y}_{i_1}$ and ${\bf Y}_{i_2}$ matrices as Y components of frames $i_1$ and $i_2$, respectively, $\sigma_{i_1,i_2}$ is calculated as $10\log_{10}(255^2/MSE({\bf Y}_{i_1},{\bf Y}_{i_2}))$; function $MSE(\cdot)$ measures the mean square error between ${\bf Y}_{i_1}$ and ${\bf Y}_{i_2}$.


Having obtained the PSNR values $\sigma_{i_1,i_2}$, we then calculate the average PSNR of each GOP, if only $\ell$ layers ($0\leq \ell\leq L$) are decodable. To this end, the frames of the undecodable layers of the current GOP are replaced by the nearest frames (in time) of decodable layers of current or previous GOPs. This is to conceal the errors. Therefore, the average PSNR of the $g$-th GOP, denoted by $\overline{\sigma_g}$, is obtained as:
\begin{align}
\overline{\sigma_g} = \Big(\sum_{i\in\mathcal{D}}\sigma_{i,i}+\sum_{i\notin\mathcal{D}}\sigma_{i,j(i)}\Big)/8
\end{align}
where the set $\mathcal{D}$ holds the indices of the frames of the decocable layers of the $g$-th GOP and $j(i)$ represents the index of the nearest decodable frame to the $i$-th frame. 

To make the PSNR calculation more clear, let us consider the following example.

\begin{exmp}
We consider a $4$-layer case, as shown in Fig.~\ref{Fig_GOP}, and assume that for the $g$-th GOP, the $3$rd layer (and consequently the $4$th one) is lost. If the I frame of the $(g-1)$-th GOP is decoded, the error concealment can be considered as shown in Fig.~\ref{Fig_GOP2} and the average PSNR is acquired as:
\ifCLASSOPTIONonecolumn
\begin{align} \label{avg_PSNR_example}
\overline{\sigma_g} = (\sigma_{k,k-1}+\sigma_{k+1,k+3}+\sigma_{k+2,k+3}+\sigma_{k+3,k+3}+\sigma_{k+4,k+3}+\sigma_{k+5,k+7}+\sigma_{k+6,k+7}+\sigma_{k+7,k+7})/8
\end{align}
\else
\begin{align} \label{avg_PSNR_example}
\overline{\sigma_g} = (&\sigma_{k,k-1}+\sigma_{k+1,k+3}+\sigma_{k+2,k+3}+\sigma_{k+3,k+3} \nonumber \\
&+\sigma_{k+4,k+3}+\sigma_{k+5,k+7}+\sigma_{k+6,k+7}+\sigma_{k+7,k+7})/8
\end{align}
\fi
If the I frame of the $(g-1)$-th GOP is not decoded, the $(k+3)$-th frame is displayed to conceal the loss of the $k$-th frame and the first term in~\eqref{avg_PSNR_example} needs to be replaced with $\sigma_{k,k+3}$.

\begin{figure}[!h]
\centering
\includegraphics[width=2.3in]{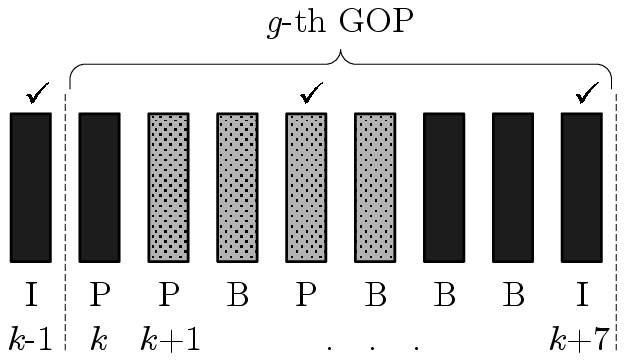}
\caption{An example of using the nearest decoded frames to conceal the loss of undecoded frames.} \label{Fig_GOP2}
\end{figure}
\vspace{-5mm}
\qed
\end{exmp}

We note that in practice the calculation of $\sigma_{i_1,i_2}$ can be done at the sender side when video is encoded. Then to obtain the PSNR performance, $\overline{\sigma_g}$ can be calculated at either the sender or the user side. For the former, the reception of number of layers for each GOP can be sent back to the sender infrequently via the available backward link for control signals shown in Fig.~\ref{Sys_model}. For the latter, the sender can embed the information into packet header, e.g., RTP\footnote{Real-time Transport Protocol, a protocol for delivering audio and video}~\cite{Schulzrinne:2003:RTP:RFC3550} extended header or part of the network coding header, as explained in~\cite{Hulya:2009:IEEE-JSAC:VAO} so that the user can calculate $\overline{\sigma_g}$.

\section{Numerical Results}
\label{RESULTS}

\begin{figure}[t]
\centering
\includegraphics[width=3.3in]{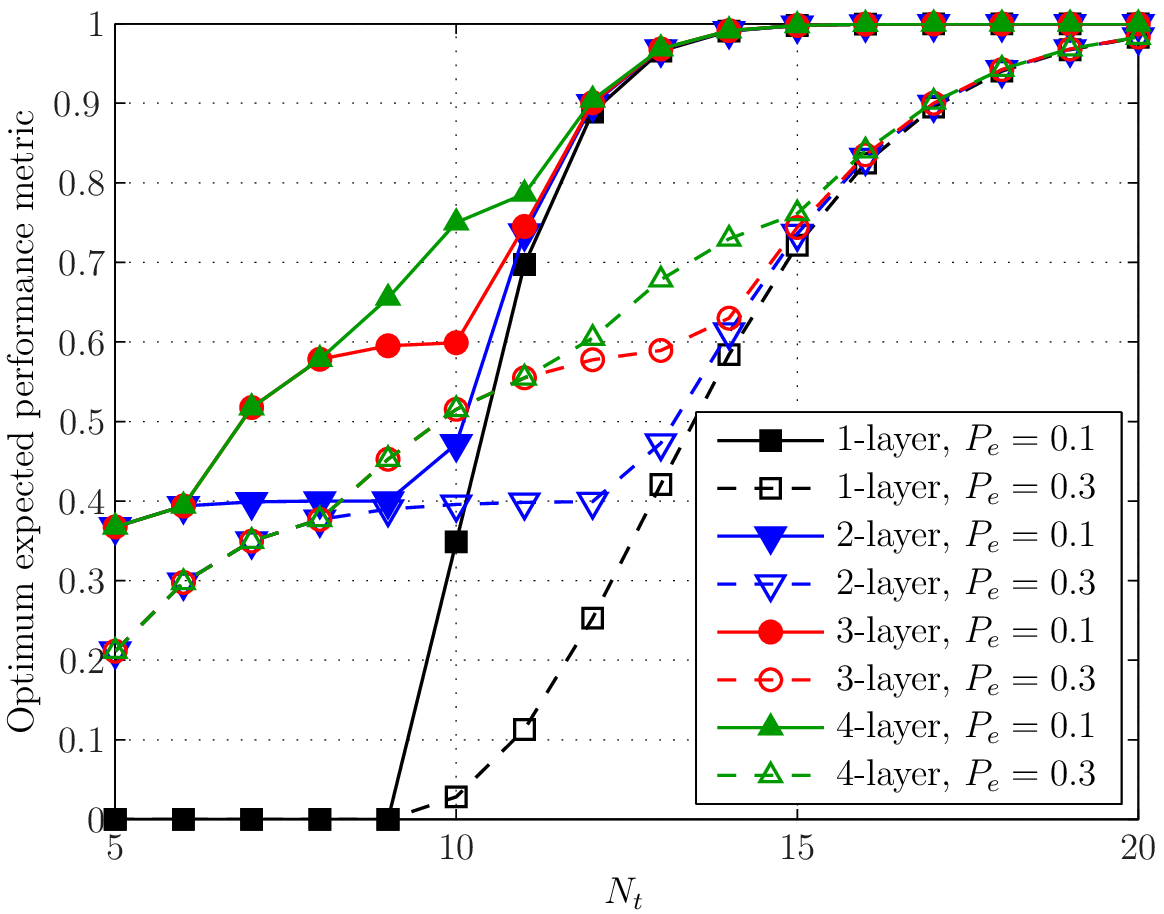}
\caption{Optimum expected performance metric versus $N_t$ for $M=10$ packets using the feedback-free scheme. Different number of layers and consequently different number of packets per layer are considered as follows: $1$-layer with ${\bf K}=[10]$, $2$-layer with ${\bf K}=[4, 6]$, $3$-layer with ${\bf K}=[4, 2, 4]$ and $4$-layer with ${\bf K}=[4, 2, 2, 2]$.} \label{Fig3}
\end{figure} 

In this section, we present the numerical results in three main parts. First, we discuss a simple case for a layered data and study the effect of the number of layers on the theoretical performance metrics. Second, we consider real video streams and discuss the theoretical performance metric and the PSNR performance for the single-user case and finally extend the discussion to the multi-user case. In each of these parts, the feedback-free scheme is the main focus and we compare its performance with the full-feedback scheme as well as an uncoded scheme, when appropriate. The details of the uncoded scheme and formulations of the theoretical performance metric are provided in Appendix~\ref{Appendix_Uncoded}.

Results in the first part are the evaluation outcomes of the analytical expressions in Section~\ref{feedback_free_1user} and Appendix~\ref{Appendix_Uncoded}. In the second and the third parts, the analytical expressions in Sections~\ref{Formulation_single},~\ref{Formulation_multi} and Appendix~\ref{Appendix_Uncoded} are used for the system design and then simulation results are obtained and shown. All the implementations and simulations are done in MATLAB\footnote{We have benefited from mex-files to expedite the simulations and also MATLAB sparse matrix functionalities to reduce the required memory.}. An MDP toolbox~\cite{chades2012markov} is employed for the finite horizon MDP modeling used in the full-feedback scheme. The optimizations in~\eqref{single_opt_feedbackfree} and~\eqref{Free_Multi2} are solved by exhaustively searching through all possible cases.

\subsection{Theoretical performance metric -- a simple example}

We consider a layered data, which has $M=10$ packets. We consider up to $4$ layers and without loss of generality, ignore the effect of number of layers on $M$, mentioned in Remark~\ref{remark3}. Hence, we assume $4$ different cases with different number of layers, ${\bf K}=[10]$, ${\bf K}=[4, 6]$, ${\bf K}=[4, 2, 4]$ and ${\bf K}=[4, 2, 2, 2]$ and calculate the optimum performance metric for the feedback-free scheme by using~\eqref{single_opt_feedbackfree} with the coefficients defined in~\eqref{coef_throughput}. The obtained results for two choices of PER, $P_e=0.1$ and $P_e=0.3$, are depicted in Fig.~\ref{Fig3}. Each point in this plot corresponds to an optimal ${\bf {N^T}^*}$, obtained based on $N_t$, $P_e$ and $L$.

Considering the results shown, it can be observed that using more layers clearly provides a better performance. However, this improved performance is only for an intermediate range of $N_t$ values, where the number of transmissions is large enough to deliver a subset of packets, but not enough to deliver all the $M$ packets. It can be seen, and also easily inferred, that this range of $N_t$ is dependent upon the PER values, as higher PER values require more transmissions to provide a specific performance, thus the mentioned range of improvement shifts to the higher values of $N_t$ as PER increases. 

\begin{figure}[!t]
\centering
\includegraphics[width=3.3in]{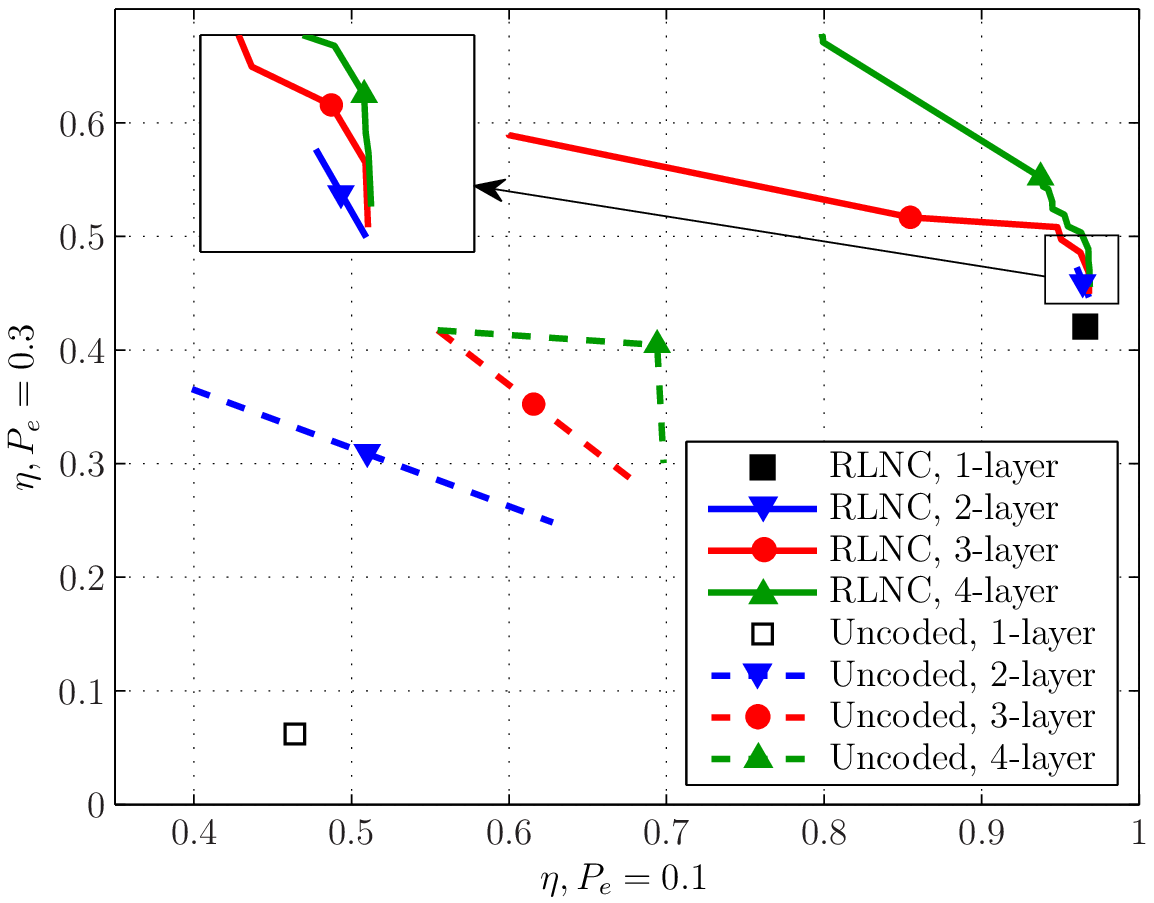}
\caption{Throughput performance of the feedback-free RLNC and uncoded schemes for $N_t=13$. The number of points in some curves is naturally limited; for example when L=1, there is only one point on the curve. Each point represents one Pareto optimal policy for the $2$-user case. Parameters $M$ and ${\bf K}$ are similar to those used in Fig.~\ref{Fig3}.} \label{Fig4}
\end{figure}

Note that for a chosen $N_t$, since the optimum policies ${\bf {N^T}^*}$ for two users, one with $P_e=0.1$ and the other with $P_e=0.3$, are different, the shown performances cannot be obtained simultaneously for the multi-user cases. In other words, the optimum policy for a user with $P_e=0.1$ will result in a worse performance than the one shown in Fig.~\ref{Fig3} for a user with $P_e=0.3$. Hence, for the multi-user case, there is a trade-off among the performances of different users and an absolute optimum ${\bf{N^T}^*}$ may not be found. Therefore, \emph{Pareto} optimal~\cite{Sawaragi:1985:TMO} ${\bf {N^T}^*}$ for multi-user case should be considered. This is shown in Fig.~\ref{Fig4} for a $2$-user case using $N_t=13$.

\ifCLASSOPTIONtwocolumn
\begin{figure*}
\centering
\includegraphics[width=6.1in]{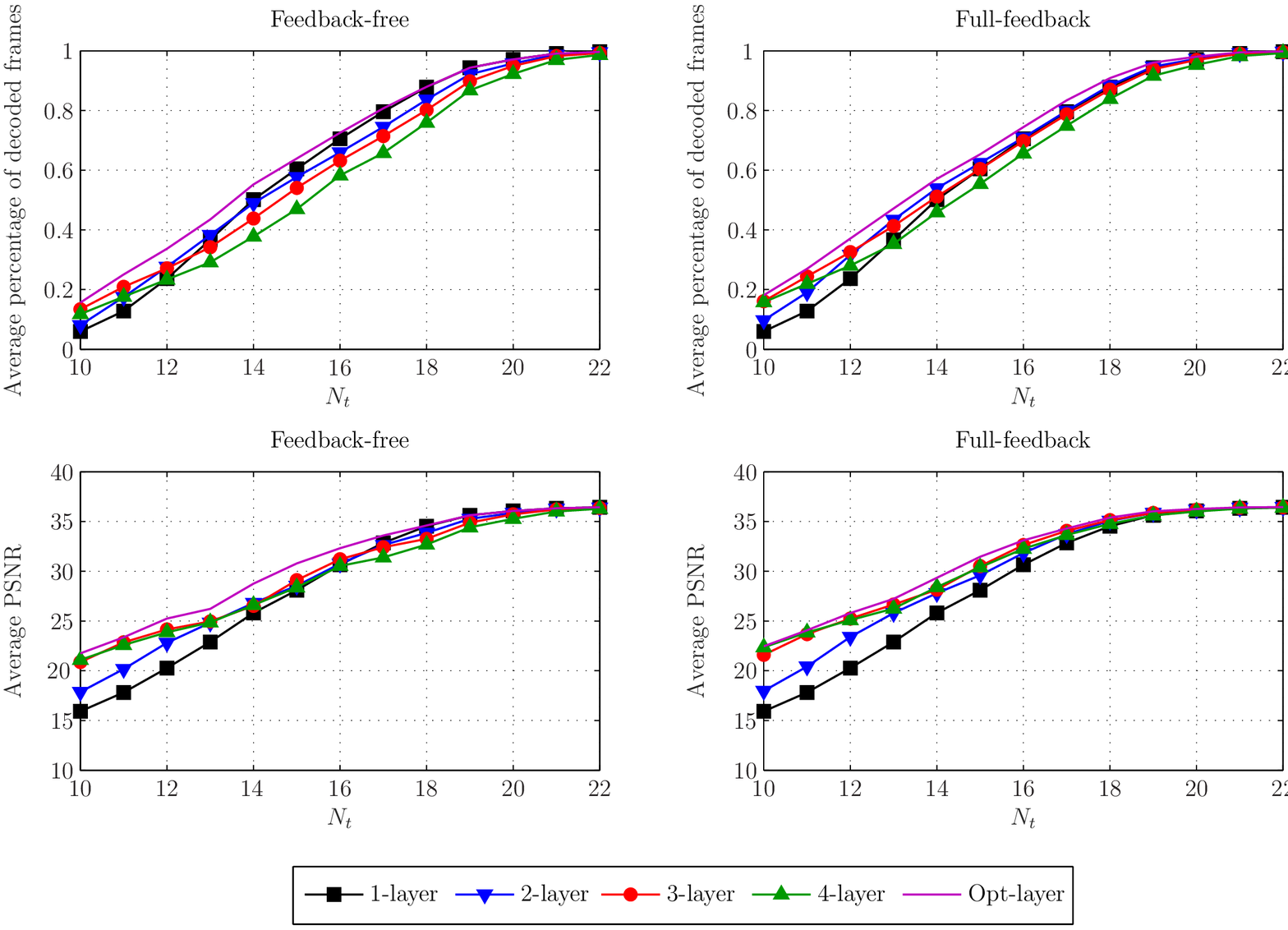}
\caption{Average percentage of decoded frames and average PSNR for single-user case using \emph{Foreman} test stream. Feedback-free and full-feedback schemes with different number of layers are compared. Values are averaged over the 38 GOPs and the 100 repetitions. } \label{Fig5}
\end{figure*}
\fi

The results in Fig.~\ref{Fig4} show that as the number of layers are increased, more Pareto optimal policies are possible, which means there are more opportunities to optimize and design the multi-user broadcasting system. This, in fact, implies that using more layers results in a better trade-off between the performances of users. However, similar to the results in Fig.~\ref{Fig3}, the gain of using more layers is not for every choice of $N_t$. For instance, for $N_t=20$ that is large enough to deliver all the $M=10$ packets, using $4$ layers theoretically does not provide any advantage over a single layer case. Moreover, we should remember that this gain is obtained by ignoring the effect of number of layers on total number of packets, discussed in Remark~\ref{remark3}. In the next subsections, we will consider this remark and observe its effect on the results.

Fig.~\ref{Fig4} also compares the feedback-free RLNC scheme with the feedback-free uncoded scheme. It can be observed that the former significantly outperforms the latter.
  
\subsection{Results for real video streams -- Single-user case} \label{real_single}

\ifCLASSOPTIONtwocolumn
\begin{table*} 
\renewcommand{\arraystretch}{1.35}
\caption{Maximum improvement (in the average performance metrics) of full-feedback scheme over feedback-free scheme for various test streams and test setups. Means are also provided in parentheses.}
\label{Table_result}
\centering
\begin{tabular}{|cc||cccc|cccc|}
\hline
& & \multicolumn{4}{c|} {Max. (mean) improvement in theoretical $\eta$} &
\multicolumn{4}{c|}{Max. (mean) improvement in PSNR (dB)}\\
               & $P_e$& $2$-layer & $3$-layer & $4$-layer & opt-layer& $2$-layer & $3$-layer & $4$-layer& opt-layer   \\
\hline
\multirow{2}{*}{Foreman}&$0.1$ & $5.3~(2.0)~\%$ & $7.3~(2.9)~\%$ & $9.1~(3.3)~\%$ & $3.7~(1.2)~\%$& $1.21~(0.40)$& $1.88~(0.67)$& $2.25~(0.83)$ & $1.00~(0.34)$    \\
&$0.3$ & $8.4~(4.0)~\%$ & $11.9~(6.1)~\%$ & $13.1~(7.0)~\%$ & $6.3~(2.9)~\%$ &$2.22~(0.90)$&$3.00~(1.52)$&$3.53~(1.93)$&$2.22~(1.04)$    \\
\hline
\multirow{2}{*}{Crew}&$0.1$ & $6.7~(2.6)~\%$ & $7.8~(3.7)~\%$ & $7.9~(4.1)~\%$ & $4.5~(2.3)~\%$ &$1.01~(0.42)$&$1.24~(0.65)$&$1.22~(0.78)$&$1.17~(0.47)$   \\
&$0.3$ & $9.1~(4.6)~\%$ & $11.6~(6.8)~\%$ & $11.9~(7.7)~\%$ & $7.7~(5.3)~\%$ &$1.52~(0.73)$&$2.10~(1.23)$&$2.12~(1.52)$&$1.74~(1.15)$    \\
\hline
\multirow{2}{*}{Soccer}&$0.1$ & $5.8~(2.6)~\%$ & $7.0~(3.6)~\%$ & $7.2~(4.1)~\%$ & $5.0~(2.4)~\%$ &$1.01~(0.52)$&$1.59~(0.80)$&$1.96~(0.95)$&$1.51~(0.57)$    \\
&$0.3$ & $8.5~(4.7)~\%$ & $11.0~(7.0)~\%$ & $12.2~(8.0)~\%$ & $8.1~(5.4)~\%$  &$1.73~(0.97)$&$2.52~(1.56)$&$2.88~(1.91)$&$2.44~(1.40)$  \\
\hline
\end{tabular}
\end{table*}
\fi

In this subsection, we consider real video test streams and in contrast to the previous subsection, we take into account the effect of number of layers $L$ on the total number of packets $M$, discussed in Remark~\ref{remark3}. The purpose is to investigate the performance (i.e., the average percentage of decoded frames and PSNR) of the proposed feedback-free RLNC scheme in comparison to the idealistic full-feedback scheme. To this end, for each GOP with known number of packets per layer (${\bf K}$), we first design the optimum transmission policy, i.e., obtain the ${\bf {N^T}^*}$ for the feedback-free scheme and $\pi({\bf s},t)$ for the full-feedback scheme, based on the values of $N_t$, $L$, $P_e$ and the calculated theoretical performance metric. Then, we model the erasure channels and assess how the designed policies perform under simulations.

To model the erasure channel, we consider time is slotted and each slot is only enough for transmission of one packet of $1500$ bytes. We assume that the channel is either in ON state, with probability $1-P_e$, or in OFF state, with probability $P_e$. Hence, we generate random erasure patterns of $38\times N_t$ long\footnote{Note that each of the test streams is composed of 38 GOPs, as mentioned in Section~\ref{SVC_PSNR}.} containing $0$ and $1$ elements, where $0$ and $1$ correspond to the OFF and ON states of the channel, respectively. 

Given an erasure pattern and the optimum transmission policies, it is straightforward to obtain the highest decodable layer of both the feedback-free and full-feedback schemes for any GOP, by using~\eqref{feedback-free_LMax} and~\eqref{transition_func_full}, respectively. Then, we calculate the performance metrics with regard to the importance of decoded layers mentioned in Section~\ref{SVC_PSNR}. We repeat this for $100$ different erasure patterns and calculate the average performance metrics.

\ifCLASSOPTIONonecolumn
\begin{figure*}
\centering
\includegraphics[width=6.1in]{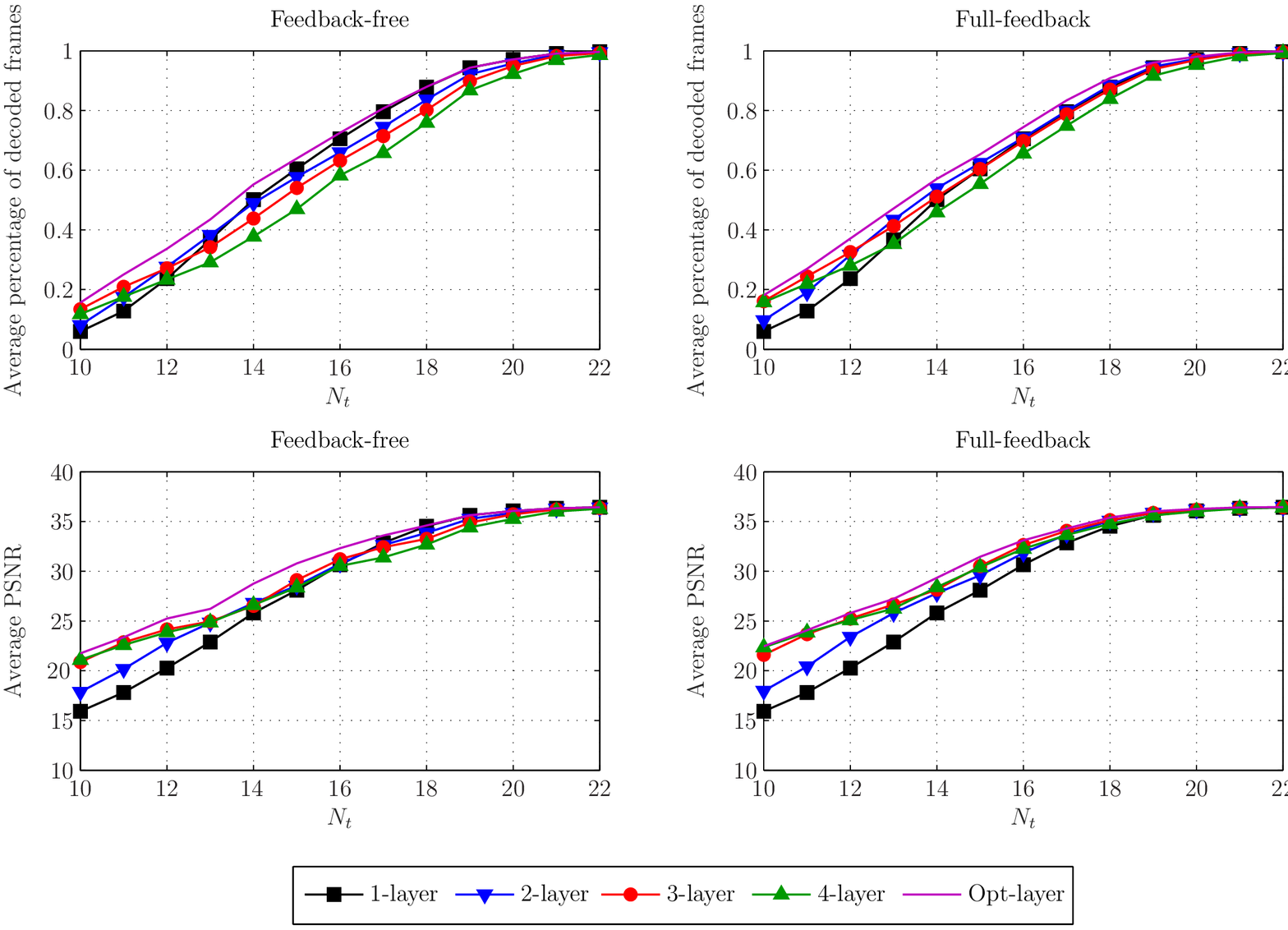}
\caption{Average percentage of decoded frames and average PSNR for single-user case using \emph{Foreman} test stream. Feedback-free and full-feedback schemes with different number of layers are compared. Values are averaged over the 38 GOPs and the 100 repetitions.} \label{Fig5}
\end{figure*}
\fi

The results in Fig.~\ref{Fig5} illustrate the average performance metrics for both schemes for the~\emph{Foreman} test stream. In addition, Table~\ref{Table_result} highlights the maximum and mean improvement (in the average performance metrics) of the full-feedback scheme over the feedback-free scheme for all the three test streams, where the maximum and mean values are calculated across $10 \leq N_t \leq 30$. Note that the $1$-layer case performs similarly in both schemes, as there is only one type of coded packet for transmission (i.e., only one action in the finite horizon MDP), so feedback cannot provide any extra gain. Therefore, the graphs for the $1$-layer case can be used as a reference to compare graphs of feedback-free and full-feedback schemes and it is not presented in Table~\ref{Table_result} as the improvement is always zero. Here, the main observations can be summarized as follows:

\ifCLASSOPTIONonecolumn
\begin{table*} 
\renewcommand{\arraystretch}{1.25}
\caption{Maximum improvement (in the average performance metrics) of full-feedback scheme over feedback-free scheme for various test streams and test setups. Means are also provided in parentheses.}
\label{Table_result}
\centering
\resizebox{1\textwidth}{!}{
\begin{tabular}{|cc||cccc|cccc|}
\hline
& & \multicolumn{4}{c|} {Max. (mean) improvement in theoretical $\eta$} &
\multicolumn{4}{c|}{Max. (mean) improvement in PSNR (dB)}\\
               & $P_e$& $2$-layer & $3$-layer & $4$-layer & opt-layer& $2$-layer & $3$-layer & $4$-layer& opt-layer   \\
\hline
\multirow{2}{*}{Foreman}&$0.1$ & $5.3~(2.0)~\%$ & $7.3~(2.9)~\%$ & $9.1~(3.3)~\%$ & $3.7~(1.2)~\%$& $1.21~(0.40)$& $1.88~(0.67)$& $2.25~(0.83)$ & $1.00~(0.34)$    \\
&$0.3$ & $8.4~(4.0)~\%$ & $11.9~(6.1)~\%$ & $13.1~(7.0)~\%$ & $6.3~(2.9)~\%$ &$2.22~(0.90)$&$3.00~(1.52)$&$3.53~(1.93)$&$2.22~(1.04)$    \\
\hline
\multirow{2}{*}{Crew}&$0.1$ & $6.7~(2.6)~\%$ & $7.8~(3.7)~\%$ & $7.9~(4.1)~\%$ & $4.5~(2.3)~\%$ &$1.01~(0.42)$&$1.24~(0.65)$&$1.22~(0.78)$&$1.17~(0.47)$   \\
&$0.3$ & $9.1~(4.6)~\%$ & $11.6~(6.8)~\%$ & $11.9~(7.7)~\%$ & $7.7~(5.3)~\%$ &$1.52~(0.73)$&$2.10~(1.23)$&$2.12~(1.52)$&$1.74~(1.15)$    \\
\hline
\multirow{2}{*}{Soccer}&$0.1$ & $5.8~(2.6)~\%$ & $7.0~(3.6)~\%$ & $7.2~(4.1)~\%$ & $5.0~(2.4)~\%$ &$1.01~(0.52)$&$1.59~(0.80)$&$1.96~(0.95)$&$1.51~(0.57)$    \\
&$0.3$ & $8.5~(4.7)~\%$ & $11.0~(7.0)~\%$ & $12.2~(8.0)~\%$ & $8.1~(5.4)~\%$  &$1.73~(0.97)$&$2.52~(1.56)$&$2.88~(1.91)$&$2.44~(1.40)$  \\
\hline
\end{tabular}}
\end{table*}
\fi

\begin{itemize}
\item for the feedback-free scheme, it can be observed that increasing the number of layers $L$ does not always improve the performance. However, this is not surprising as, according to Remark~\ref{remark3}, by increasing $L$, which enables decoding of a subset of a GOP's frames to improve the performance, the total number of packets $M$ may also increase, which can inversely affect the performance. 
\item comparing the feedback-free and full-feedback schemes, Fig.~\ref{Fig5} reveals that the feedback-free scheme can perform very close to the full-feedback scheme for the considered \emph{Foreman} stream. This is confirmed by the detailed results in Table~\ref{Table_result}, which show the full-feedback scheme can only improve the PSNR by $1\sim2$ dB and $2\sim3$ dB for users with PER of $0.1$ and $0.3$, respectively.
\item considering the opt-layer approach, it is evident in Fig.~\ref{Fig5} that it outperforms all other approaches with fixed number of layers in both feedback-free and full-feedback schemes. Moreover, as Table~\ref{Table_result} suggests, selecting the number of layers for each GOP adaptively results in performances that are theoretically closer to those of an idealistic full-feedback scheme. 
\item while the general behaviors/trends of the average percentage of decoded frames and the average PSNR are similar with respect to $N_t$, the effect of different number of layers for some values of $N_t$ are different. This is due to the fact that in calculating $\eta$, a contribution of zero is considered to the performance metric if a layer is not decoded, whereas for the PSNR, error concealment, i.e., showing a decoded frame instead of a lost frame, contributes to the PSNR.  
\end{itemize}

\subsection{Results for real video streams -- Multi-user case} \label{real_Multi}

In this subsection, we assess the performance of the proposed feedback-free RLNC scheme for the multi-user case. The assessment setup is similar to the previous subsection, i.e., the schemes to be tested are first designed for each GOP and are then tested under some simulated erasure patterns. However, it is required that an aggregate performance function $H(\cdot)$ is specified. Here, we first consider a general linear function to combine performances of users and compare the feedback-free and full-feedback schemes. Then, we focus on the mean and fairness~\cite{Fairness_jain} of users' performances as more specific aggregate functions and combine them to show the performance trade-offs. \emph{Foreman} test stream is used throughout this subsection.

\subsubsection{General case}

As mentioned previously, to design a transmission scheme for multiple users, various aggregate performance metrics are possible. Here, in order to make a thorough comparison between the feedback-free RLNC and idealistic full-feedback schemes, we consider a series of aggregate functions, i.e., we define $H(\cdot)$ as follows:
\begin{align} \label{aggregate_func}
H(z_1,..., z_{N_u}) = \sum_{i=1}^{N_u}w_iz_i
\end{align}
and use various weight vectors ${\bf W}=[w_1, ..., w_{N}]$, such that $\sum_{i=1}^{N_u}w_i=1$. The input argument $z_i$ is replaced by $\eta_i$ and $G_{\text{single}}({\bf s_i})$ for feedback-free and full-feedback schemes, respectively.

Due to the computational complexities of MDP for very large state spaces, we limit our study in this section to $N_u=3$ users and maximum $L=3$ layers. Packet error rates are assumed to be $P_{e_1}=0.1$, $P_{e_2}=0.15$ and $P_{e_3}=0.2$ and the weights $w_i$ are chosen from $\{0,1/3,2/3,1\}$, which result in 10 unique weight vectors. We obtain the optimal policies by using~\eqref{Free_Multi1},~\eqref{Free_Multi2} and~\eqref{MDP_Multi} for each GOP and different $N_t$ values and then similar to previous subsections, test the designed policies for random erasure patterns corresponding to each user. This process is repeated $100$ times for each of the $10$ weight vectors. Moreover, fixed number of layers ($1$, $2$ and $3$) and also optimum number of layers are considered.

\begin{figure*}
\centering
\includegraphics[width=6.3in]{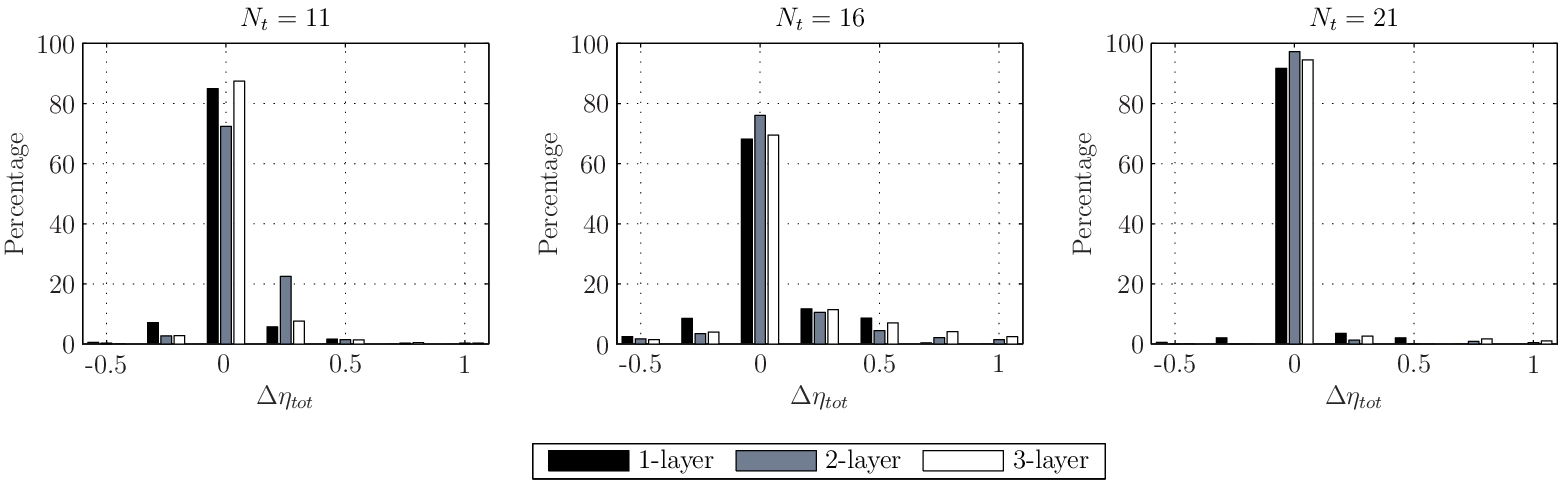}
\caption{Histograms showing the difference of $\eta_{tot}$ between feedback-free schemes with optimum number of layers and with fixed number of layers for a $3$-user case using \emph{Foreman} test stream. Positive $\Delta\eta_{tot}$ values correspond to higher $\eta_{tot}$ values of the opt-layer scheme.} \label{Fig6}
\end{figure*}

The first results, depicted in Fig.~\ref{Fig6}, reveal the effect of choosing the number of layers adaptively for each GOP. The histograms show in what percentage of the cases opt-layer scheme improves the theoretical performance metric for the feedback-free RLNC scheme. $\Delta \eta_{tot}$ in $x$-axis is defined as the difference of $\eta_{tot}$ between the opt-layer and fixed layer schemes, where positive values correspond to higher $\eta_{tot}$ of the opt-layer scheme. 

\ifCLASSOPTIONtwocolumn 
\begin{figure*}
\centering
\includegraphics[width=6.3in]{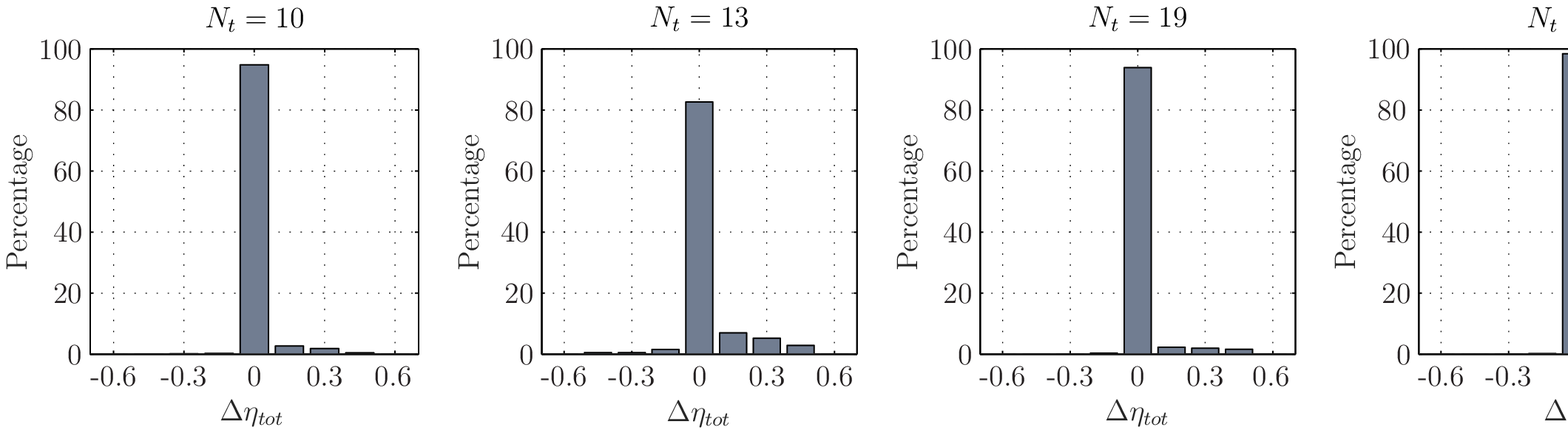}
\caption{Histograms showing the difference of $\eta_{tot}$ between feedback-free and idealistic full-feedback schemes with optimum number of layers for a $3$-user case using \emph{Foreman} test stream. Positive $\Delta\eta_{tot}$ values correspond to higher $\eta_{tot}$ values of idealistic full-feedback scheme.} \label{Fig7}
\end{figure*}
\fi

It is observed that the proposed opt-layer scheme that selects the optimum number of layers for each GOP based on the analytical results, outperforms the schemes with fixed number of layers in simulation. Similar to the results for the single-user case, the amount of improvement varies based on $N_t$ and also $L$, and diminishes as $N_t$ gets larger.

Next, we consider the opt-layer approach and provide the results for the comparison of $\eta_{tot}$ values between the feedback-free and the idealistic full-feedback schemes that are depicted in Fig.~\ref{Fig7}. We use histograms again and $\Delta \eta_{tot}$ for $x$-axis, where positive values of $\Delta \eta_{tot}$ correspond to higher $\eta_{tot}$ of idealistic full-feedback schemes. (Note that the depicted range of values for $\Delta \eta_{tot}$ are different in Figs.~\ref{Fig6} and~\ref{Fig7}.)

\ifCLASSOPTIONonecolumn 
\begin{figure*}
\centering
\includegraphics[width=6.3in]{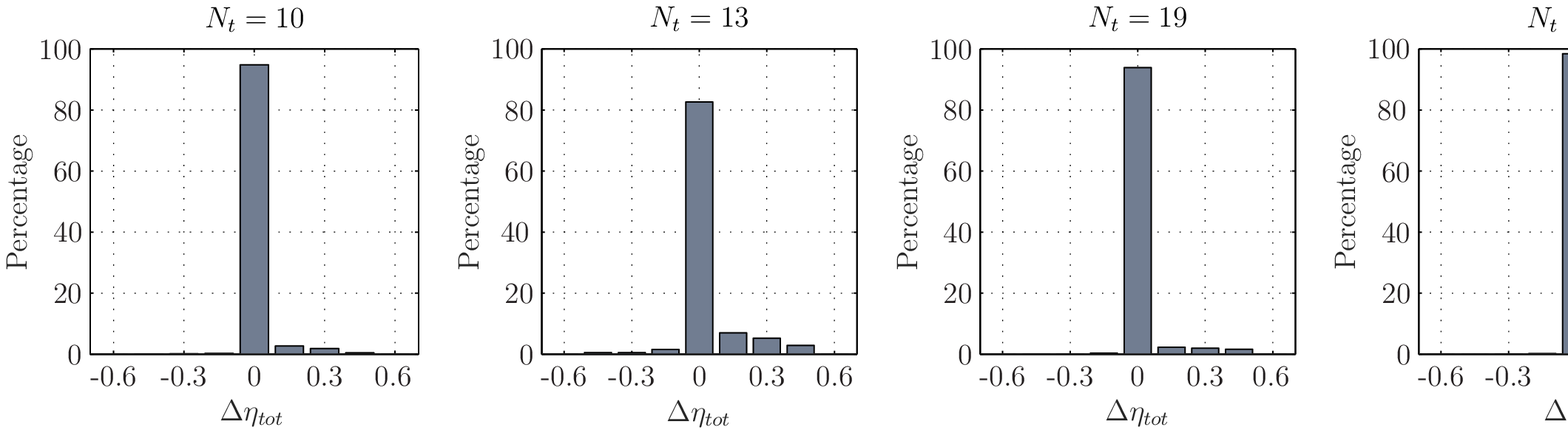}
\caption{Histograms showing the difference of $\eta_{tot}$ between feedback-free and idealistic full-feedback schemes with optimum number of layers for a $3$-user case using \emph{Foreman} test stream. Positive $\Delta\eta_{tot}$ values correspond to higher $\eta_{tot}$ values of idealistic full-feedback scheme.} \label{Fig7}
\end{figure*}
\fi

From the results, it is interesting to note that even in the worst case of $N_t=13$, the feedback-free scheme works very similar to the idealistic scheme in more than $80\%$ of the cases. As the possible number of transmissions $N_t$ increases, the two schemes work even closer. The reason behind this is understandable from the results in Fig.~\ref{Fig5}, where for large $N_t$ values, the opt-layer  is in fact the $1$-layer approach, which is identical for the feedback-free and full-feedback schemes.

\begin{figure}
\centering
\includegraphics[width=3.3in]{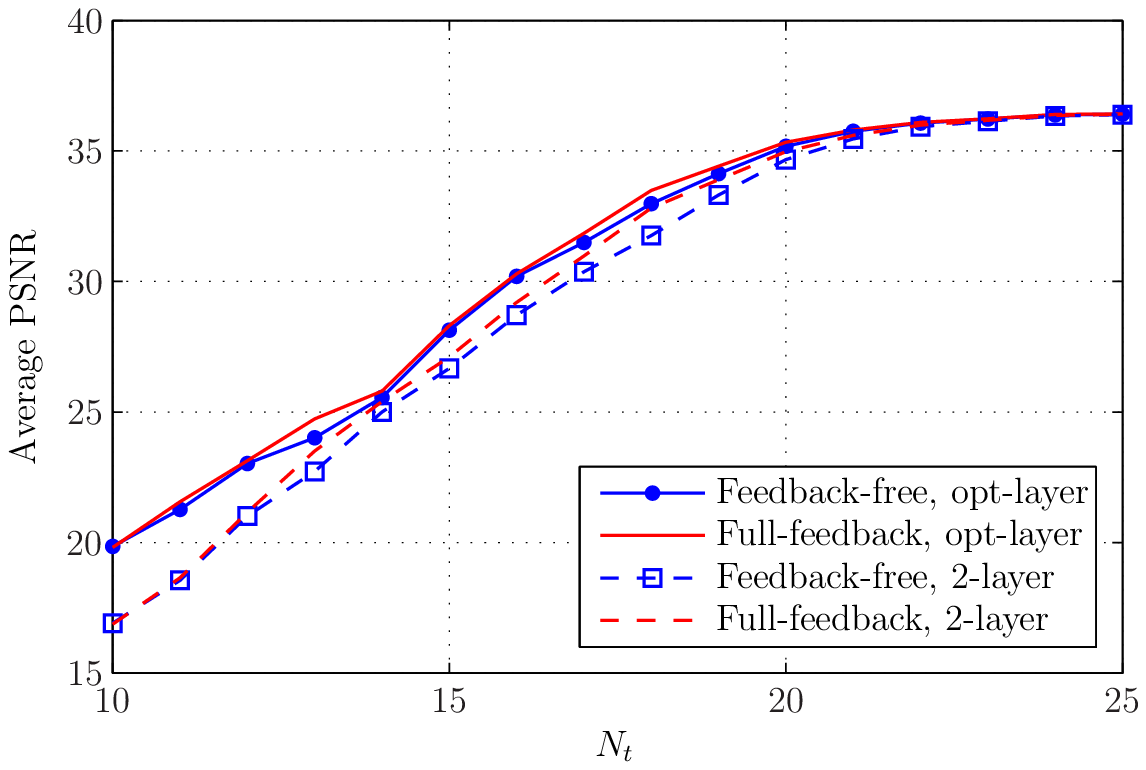}
\caption{Average PSNR for the multi-user feedback-free and idealistic full-feedback schemes. Results for opt-layer and $2$-layer cases are shown. PSNR values are averaged over the $38$ GOPs, the $100$ repetitions and the $3$ users. } \label{Fig8}
\end{figure}

In addition to the theoretical performance metric, we also compare the PSNR of different schemes. To this end, we consider the simulation setup mentioned above, but focus on a case where users have equal importance, i.e $w_1=w_2=w_3=1/3$, which means the transmission policies are designed based on the mean theoretical performance metric. We test the designed policy for each GOP and for each user and repeat this $100$ times to obtain the average PSNR, which are shown in Fig.~\ref{Fig8}.

The results confirm that the opt-layer feedback-free scheme works very close to the idealistic full-feedback scheme. In fact, by using adaptive selection of number of layers for each GOP, not only the performance of the feedback-free scheme is improved, but also the gap with the idealistic scheme reduces. For the $2$-layer and $3$-layer cases, the maximum difference in the average PSNR was $1.06$ and $1.29$ dB, which is reduced to $0.72$ dB with adaptive selection of number of layers.
%
\ifCLASSOPTIONtwocolumn 
\begin{figure*}
\centering
\includegraphics[width=6in]{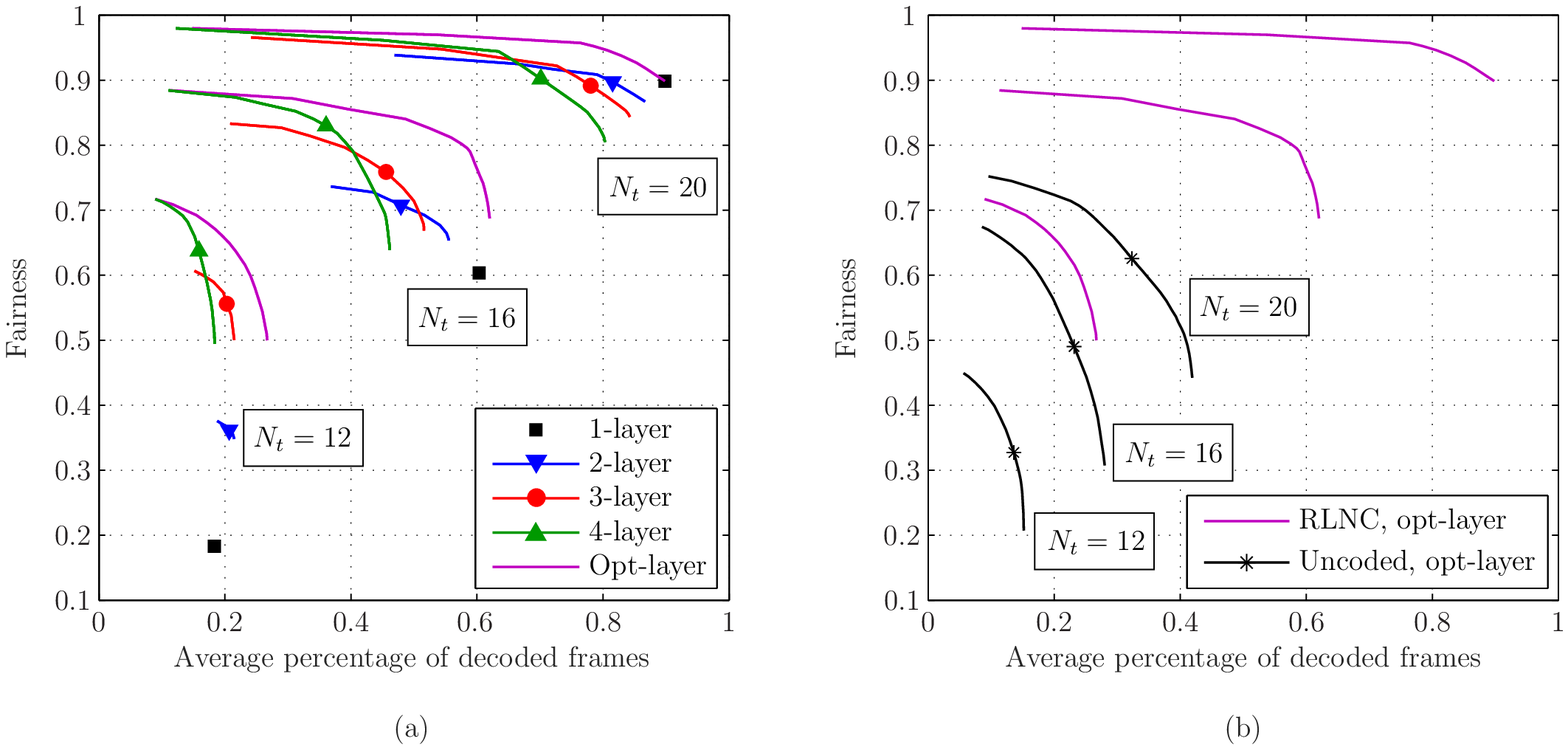}
\caption{Trade-off curves, (a) highlighting the advantage of opt-layer approach for the feedback-free RLNC scheme (b) comparing the feedback-free RLNC and uncoded schemes. Results for three values of $N_t$ are provided.} \label{Fig9}
\end{figure*}
\fi  

\subsubsection{Performance Trade-offs}

In this subsection, the purpose is to highlight the trade-offs between the performances of users with different channel conditions for the proposed feedback-free schemes. A simple example of performance trade-off was shown in Fig.~\ref{Fig3} for a two-user case by considering only one sample GOP and the theoretical results were provided. Here, we study a more comprehensive case by considering $N_u=10$ users and real video test streams, and present the simulation results. 

With the considered number of users, it is not possible to show the performance trade-offs in $N_u$ dimensions. Instead, we consider two aggregate performance functions and study their values in $2$ dimensions. We use the function defined in~\eqref{aggregate_func} and select all the weights to be $1/N_u$, so that the mean performance is calculated. This is referred to as $H_1(\cdot)$. For the second function, we use the Jain's fairness index~\cite{Fairness_jain}, which is defined as follows:
\begin{align}
H_2(z_1,...,z_{N_u}) = \frac{(\sum_{i=1}^{N_u}z_i)^2}{N_u.\sum_{i=1}^{N_u}(z_i^2)}
\end{align} 
with values ranging from $1/N_u$ (worst case: only one of the input arguments is nonzero) to $1$ (best case: all the input arguments are equal). Considering $\eta_i$'s as the input arguments, maximizing this fairness index leads to designing a fair policy in the sense that all users achieve similar theoretical performances, regardless of their channel conditions.

Having defined the two aggregate functions, the problem of optimizing their values is a bi-objective optimization. Since, an optimal solution (i.e., a feedback-free transmission policy here) that can maximize both objectives concurrently may not be found in all cases, we obtain the Pareto optimal solutions and study the trade-off between the two objectives. The most common technique to do this is the \emph{weighted sum} method~\cite{Marler:2004:SMO}, which combines the objectives as follows:
\ifCLASSOPTIONonecolumn
\begin{align}
\eta_{tot} = H(\eta_1,...,\eta_{N_u}) = \lambda H_1(\eta_1,...,\eta_{N_u}) + (1 - \lambda)H_2(\eta_1,...,\eta_{N_u})
\end{align}
\else
\begin{align}
\eta_{tot}=H(\eta_1,...,\eta_{N_u}) =& \lambda H_1(\eta_1,...,\eta_{N_u}) \nonumber \\
+&(1 - \lambda)H_2(\eta_1,...,\eta_{N_u})
\end{align}
\fi
Then using this $\eta_{tot}$ in~\eqref{Free_Multi2}, for every value of $0\leq \lambda \leq 1$, a Pareto optimal solution is obtained~\cite{Marler:2004:SMO}. Selecting $51$ values for $\lambda$ from $[0, 1]$ with step size of $0.02$, the trade-off curves shown in Fig.~\ref{Fig9} are resulted.

\ifCLASSOPTIONonecolumn 
\begin{figure*}
\centering
\includegraphics[width=6in]{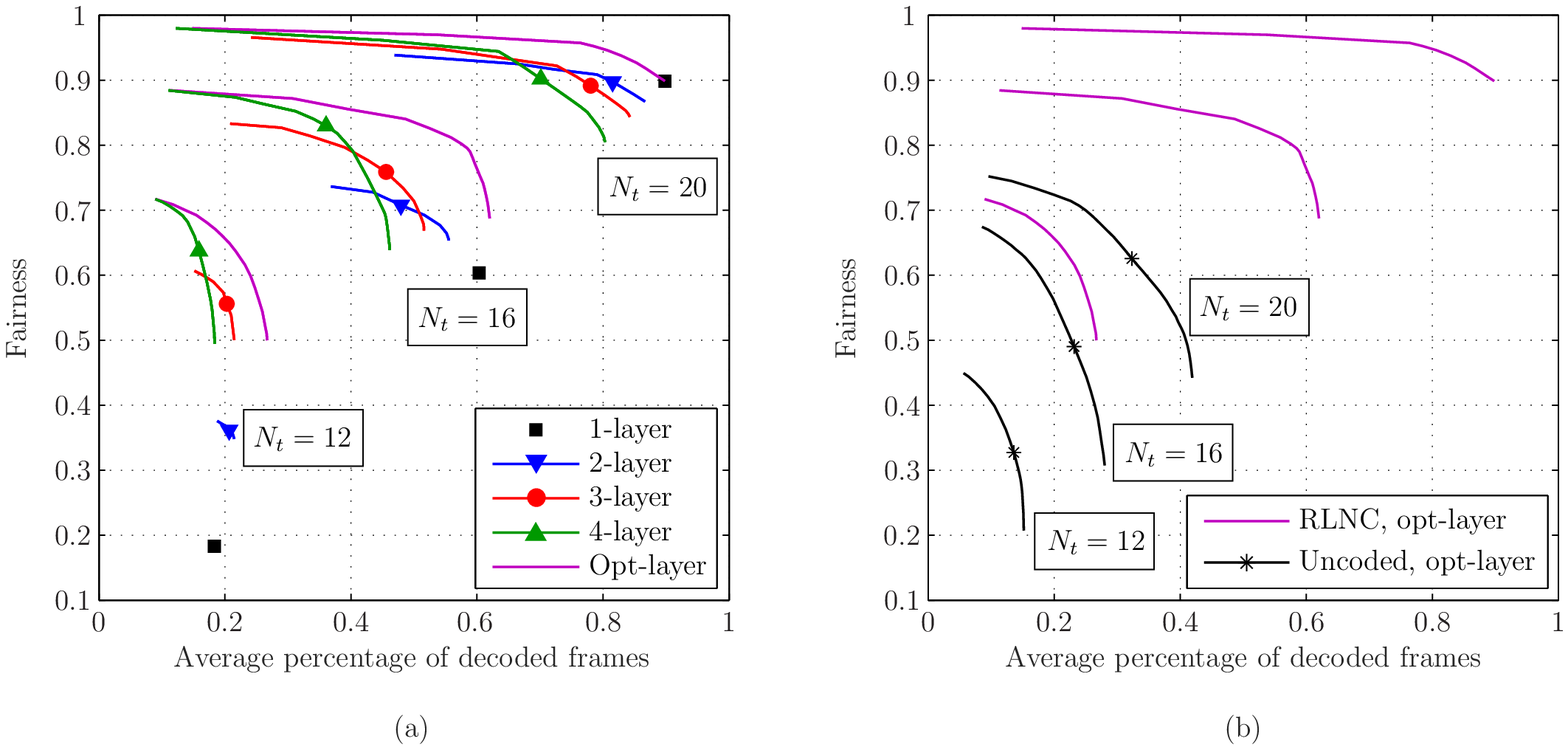}
\caption{Trade-off curves, (a) highlighting the advantage of opt-layer approach for the feedback-free RLNC scheme (b) comparing the feedback-free RLNC and uncoded schemes. Results for three values of $N_t$ are provided.} \label{Fig9}
\end{figure*}
\fi  

For the results shown, we have considered the PERs of users to be $P_{e_i} = i\times0.05$ for $1\leq i \leq 5$ and $P_{e_i} = (i-5)\times0.05$ for $6\leq i \leq 10$, and similar to the previous subsections, we have generated random erasure patterns corresponding to these PERs and then obtained the simulation results. Simulations for each GOP and each $\lambda$ are repeated $50$ times and the averages are presented. The main observations are summarized as follows:

\begin{itemize}
\item trade-offs between the average percentage of decoded frames and the fairness are evident for different system parameters. The trade-offs show that selecting the operating point based on maximizing the average percentage of decoded frames does not lead to the fairest possible policy, but a fairer policy can be chosen at the expense of reducing the average percentage of decoded frames.  
\item the advantage of selecting the number of layers adaptively over fixed layer approaches can be observed in Fig.~\ref{Fig9}(a). For instance, for $N_t=16$, fairness of $0.8$ is achieved in the opt-layer approach with average percentage of decoded frames of $0.6$, where the same fairness for the 3- and 4-layer approaches are achieved with average percentage of decoded frames of $0.4$, i.e., around $50\%$ improvement is gained.
\item Fig.~\ref{Fig9}(b) clearly shows that the feedback-free RLNC outperforms the uncoded scheme.
\end{itemize}

Note that for the 1-layer case, there is only one option for transmission, i.e., coded packets from the first layer, hence all the different values of $\lambda$ resulted in the single point, which is shown in Fig.~\ref{Fig9}(a) for each $N_t$.
          
\section{Conclusion and Discussion}

Network coding combination with unequal error protection has shown to be a promising choice to enhance streaming of layered video~\cite{Nguyen:2010:VSN,Thomos:2011:IEEE-TMM:PDV, Vukobratovic:2012:IEEE-TCOM:UEP, Nazir:2012:ALS:h264,Tassi:2015:IEEE-JSAC}. However, there are still some knowledge gaps before this technique can be readily applied in practice. In this study, we focused on feedback-free UEP+RLNC and studied some of the unaddressed issues for multi-user broadcasting of layered video over heterogeneous erasure channels. In particular, we contributed by investigating the benefits of layered approach for multi-user video streaming, comprehensively studying the effect of number of video layers on the broadcast system design and quantifying the performance degradation due to not using feedback. The results show that the proposed feedback-free approach that uses adaptive selection of number of video layer performs very close to the idealistic full-feedback system and thus is a promising candidate to enhance streaming of layered video in practice. 

Although we focused only on the temporal scalability in the simulations, our proposed framework in this paper is general and can be easily applied to spatial and quality scalability as well. In fact, the main assumptions are that the video is layered and the importance of each layer can be defined, e.g., by using~\eqref{coef_throughput} or in a similar way as in Remark~\ref{remark4}. 
A similar conclusion applies to the usage of the general aggregate function. While we used some basic aggregate functions, it is possible to consider more complicated aggregate functions to study our proposed layered approach for various multi-user scenarios. For instance, if users have different displays and are hence interested in different number of video layers (and consequently different spatial and/or temporal quality), it is possible to define the aggregate function such that the theoretical performance of each user is limited to what its device can support. Therefore, our study, with some minor modifications, can be used when heterogeneity in device types exists.

One future research direction is to extend the existing approach to transmission of multiple layered video streams, where some preliminary results are presented in~\cite{Esmaeilzadeh:2014:ITW}. Moreover, investigating variants of the proposed MDP-based idealistic scheme to be used in practice (e.g., by utilizing delayed information in MDP, considering partially observable MDP or studying ways to reduce the computational complexity) and considering other heterogeneity cases (e.g., different network topology types or different mobility conditions for users) are other possible future research directions.

\appendices
\numberwithin{equation}{section}
\section{Finite Horizon MDP} \label{Appendix_MDP}
A Markov decision process (MDP)~\cite{Puterman:1994:MDP} is defined with four main components as the inputs:
\begin{itemize}
\item finite state set $\mathcal{S}$,
\item finite action set $\mathcal{A}$,
\item transition probability function $P(\acute{s}|s,a)$, which shows the probability of going to state $\acute{s}\in\mathcal{S}$ after taking action $a\in\mathcal{A}$ in state $s\in\mathcal{S}$,
\item reward function $R(s,a)$, which shows the immediate reward caused by taking action $a\in\mathcal{A}$ while in state $s\in\mathcal{S}$,
\end{itemize}
and two components as the outputs:
\begin{itemize}
\item policy $\pi(s)$ that shows the optimum action for any possible state $s\in\mathcal{S}$,
\item value function $V_{\pi}(s)$, which is a measure of the expected reward accumulated by policy $\pi(s)$.
\end{itemize}

In the finite horizon MDP, the number of decision stages (also called horizons, denoted by $T$) are limited, hence the outputs depend on the number of stages-to-go, denoted by $t\leq T$, as well. Therefore the notation for the policy and value function should be updated as $\pi(s,t)$ and $V_{\pi}^t(s)$, respectively. Moreover, there is a reward at the final stage (i.e., $t=0$), which is called the terminal reward. This is another input component for finite horizon MDP and is denoted by $G(s)$, which represents the reward if the MDP is in state $s$ at the final stage. 

Having defined all the input components, it is shown that the optimal policy and value function for each state $s\in\mathcal{S}$ can be recursively obtained by the backward induction method as follows~\cite{Puterman:1994:MDP}:
\begin{align}
V_{\pi}^t(s) = \max_{a}\{R(s,a)+\sum_{\acute{s}\in\mathcal{S}}P(\acute{s}|s,a)V_{\pi}^{t-1}(\acute{s})\} \\
\pi(s,t) = \arg\max_{a}\{R(s,a)+\sum_{\acute{s}\in\mathcal{S}}P(\acute{s}|s,a)V_{\pi}^{t-1}(\acute{s})\}
\end{align}
where $1\leq t\leq T$ and $V_{\pi}^0(s) = G(s)$. This method is shown to have time complexity of $O(T|\mathcal{A}||\mathcal{S}|^2)$~\cite{Puterman:1994:MDP}.
 
\section{Uncoded scheme} \label{Appendix_Uncoded}

In this appendix, the uncoded scheme used for comparison is briefly explained and the formulation of its theoretical performance metric is provided.

We use the same assumptions and notations specified in Section~\ref{Section2} for this scheme and consider that out of total $N_t$, $n_\ell^t$ transmissions are dedicated to the $k_\ell$ packets of the $\ell$-th layer in a Round Robin manner such that $\sum_{\ell=1}^Ln_\ell^t=N_t$. Packets are transmitted uncoded and ${\bf N^T}=[n_1^t,n_2^t,...,n_L^t]$ is decided in advance, i.e., the scheme is feedback-free.

Assuming $b_\ell=\lfloor \tfrac{n_\ell^t}{k_\ell} \rfloor$, it can be easily inferred that $c_\ell=n_\ell^t-b_\ell k_\ell$ packets are transmitted $b_\ell+1$ times and the remaining $k_\ell-c_\ell$ packets are transmitted $b_\ell$ times. Therefore, the probability that all the packets of the $\ell$-th layer can be delivered successfully is expressed by
\begin{align}
p_\ell(k_\ell,n_\ell^t) = (1-P_e^{b_\ell})^{k_\ell-c_\ell}(1-P_e^{b_\ell+1})^{c_\ell}
\end{align}
 
Now, bearing in mind that packets of the $\ell$-th layer are useful only if packets of all the lower layers are received, the probability $P_\ell({\bf K},{\bf N^T})$, defined in Section~\ref{feedback_free_1user}, will be obtained as follows:
\ifCLASSOPTIONonecolumn
\begin{align}
P_\ell({\bf K},{\bf N^T}) = \left\{
\begin{array}{lll}
(1-p_{\ell+1}(k_{\ell+1},n_{\ell+1}^t))\prod_{j=1}^\ell p_j(k_j,n_j^t)&;& \ell < L \\
\prod_{j=1}^\ell p_j(k_j,n_j^t)&;& \ell =L
\end{array}\right.
\end{align}
\else
\begin{align}
P_\ell({\bf K}&,{\bf N^T}) = \nonumber \\ 
&\left\{
\begin{array}{lll}
(1-p_{\ell+1}(k_{\ell+1},n_{\ell+1}^t))\prod_{j=1}^\ell p_j(k_j,n_j^t)&;& \ell < L \\
\prod_{j=1}^\ell p_j(k_j,n_j^t)&;& \ell =L
\end{array}\right.
\end{align}
\fi

Similar to the feedback-free RLNC, the above probabilities can be used in~\eqref{feedbackfree_eta} to acquire the theoretical performance metric.

%

\ifCLASSOPTIONcaptionsoff
  \newpage
\fi

\bibliographystyle{IEEEtran}
\bibliography{IEEEabrv,Ref}

\end{document}